\input{aipcheck}

\documentclass[
    ,final            
  ]
  {aipproc}

\layoutstyle{6x9}

\usepackage{color}
\usepackage{natbib}
\usepackage{dcolumn}
\usepackage{amssymb}
\usepackage{amsmath}
\usepackage{bm}
\usepackage{amsfonts}
\usepackage{txfonts}
\usepackage{bbm}
\usepackage{bbold}

\newcommand{\ex}{\mathrm{e}}
\newcommand{\dd}{\mathrm{d}}
\newcommand{\dr}{_\mathrm{r}}
\newcommand{\dm}{_\mathrm{m}}
\newcommand{\gsim}{\gtrsim}

\def\spose#1{\hbox to 0pt{#1\hss}}

\def\lta{\mathrel{\spose{\lower 3pt\hbox{$\mathchar"218$}}
     \raise 2.0pt\hbox{$\mathchar"13C$}}}
\def\gta{\mathrel{\spose{\lower 3pt\hbox{$\mathchar"218$}}
     \raise 2.0pt\hbox{$\mathchar"13E$}}}

\def\setR{\mathbb{R}}
\def\setC{\mathbb{C}}

\def\setUni{\mathbb{1}}

\newcommand{\ie}{\textsl{i.e.~}}

\newcommand{\Hu}{\mathcal{H}}
\newcommand{\Ka}{\mathcal{K}}

\newcommand{\cs}{c_{_\mathrm{S}}}

\newcommand{\GN}{G_{_\mathrm{N}}}

\newcommand{\Mp}{M_{_\mathrm{Pl}}}

\def\beq{\begin{equation}}
\def\eeq{\end{equation}}
\def\bea{\begin{eqnarray}}
\def\eea{\end{eqnarray}}
\def\eqref{\ref}

\renewcommand{\(}{\left(}
\renewcommand{\)}{\right)}
\renewcommand{\[}{\left[}
\renewcommand{\]}{\right]}

\newcommand{\cte}{\mathrm{C}^{\mathrm{te}}}

\begin{document}

\title{Cosmological perturbation theory}


\author{Patrick Peter}{
  address={${\cal G}\setR\varepsilon\setC{\cal O}$ -- Institut
d'Astrophysique de Paris, UMR7095 CNRS, Universit\'e Pierre \& Marie Curie,
98 bis boulevard Arago, 75014 Paris, France}
}

\begin{abstract}
The purpose of these lectures is to give a pedagogical
overview of cosmological perturbation theory, following
the lectures given during the school. The topics treated are:

I -- The background

II -- Scalar/Vector/Tensor decomposition and the gauge issue

III -- The example of the tensor modes

IV -- Density fluctuations, transfer function and power spectrum

V -- Initial condition theory: quantum vacuum fluctuations

\noindent Most of the material presented here is available in many
well-written reviews or textbooks, so in order to avoid unnecessary
heavy presentation as well as to make sure I forget nobody, I will only
cite the review paper \cite{MFB92} as well as the book \cite{PPJPU}
from which most of the figures have been taken.
Useful extra information and different perspectives can be also
found in \cite{LNP2008} (in particular the review articles by A.~Linde
on inflation, J.~Martin on the quantum aspect of initial condition
and their subsequent squeezed evolution and C.~Ringeval on the
numerical evolution of perturbations). Ref. \cite{Weinberg-Cosmology}
provides a personal vision of S.~Weinberg with many original
proofs to well-known results, and \cite{Mukhanov:2005sc}
describes in more details the relevant physics for calculating
the quantities actually to be compared with the data.
Finally, all numerical figures are taken
from the {\sl Particle Data Group} \cite{pdg} whose latest update
is always available on the linked site.

\end{abstract}

\maketitle

\section{Introduction: the background}

Even though these lectures concern cosmological perturbation
theory, I felt an introduction to the background could be welcome,
would it be only to fix the notations, set the framework and make
apparent what the problems and questions are.

Cosmology is the part of physics that studies the Universe as a whole,
trying to make models of its overall evolution and its structure. As such,
it is a quite peculiar branch of physics, as by definition there is only one
Universe -- hence the name -- and it is impossible to make any experiment
on either its evolution or structure! From these considerations, we
immediately see that cosmology will be endowed with various intrinsic
limitations which I will discuss in due turn.

How do we, practically, describe cosmology? To begin with, one needs
a theoretical framework providing the evolution equations. This will be
general relativity: the Universe will be seen as a 4-dimensional manifold,
space-time, endowed with a metric $g_{\mu\nu}$ whose dynamics
follows from Einstein equations
\begin{equation}
G_{\mu\nu} \left( \equiv R_{\mu\nu} -\frac12 g_{\mu\nu} R \right)
+ \Lambda g_{\mu\nu}
= \frac{8\pi\GN}{c^4} T_{\mu\nu},
\label{Einstein}
\end{equation}
where $c$, the velocity of light in vacuum, will be set to unity in all
further calculations (along with $\hbar$ where it should have appeared
in the final section of these lectures), $\GN$ is Newton's constant, $T_{\mu\nu}$ is the stress-energy
tensor of the matter -- discussed later -- and $\Lambda$ the cosmological
constant. We know from observations that the latter  is probably
not vanishing, contrary to what was supposed until recently, but we can
however consider its influence as another matter fluid and include it
in $T_{\mu\nu}$. Therefore, one can, without lack of generality, send
$\Lambda\to 0$ in Eq. (\ref{Einstein}).

The Einstein tensor $G_{\mu\nu}\equiv R_{\mu\nu} - \frac12 g_{\mu\nu} R$
is defined in terms of the Ricci tensor $R_{\mu\nu}$ and scalar $R\equiv R^\mu_{\ \mu}
=g^{\mu\nu} R_{\mu\nu}$, the former stemming from a contraction of the
Riemann tensor through $R_{\mu\nu} \equiv R^\alpha_{\ \mu\alpha\nu}$. Finally,
the relation with the metric itself is made with the definition
\begin{equation}
R^\mu_{\ \nu\alpha\beta} \equiv \partial_\alpha\Gamma^\mu_{\ \nu\beta}
- \partial_\beta\Gamma^\mu_{\ \nu\alpha} + \Gamma^\mu_{\ \sigma\alpha}
\Gamma^\sigma_{\ \nu\beta}- \Gamma^\mu_{\ \sigma\beta}
\Gamma^\sigma_{\ \nu\alpha},
\label{Riemann}
\end{equation}
and the Christoffel symbols are given in terms of the metric by
\begin{equation}
\Gamma^\mu_{\ \alpha\beta} \equiv \frac12 g^{\mu\nu} \left(
\partial_\alpha g_{\nu\beta} + \partial_\beta g_{\nu\alpha}
-\partial_\nu g_{\alpha\beta}\right).
\label{Gammas}
\end{equation}
This completes the geometrical explanation, i.e. the left hand
side of Eq. (\ref{Einstein}). The next question, more physical in a way,
now is: what is the matter content of the Universe?
With this content well-defined, one can in principle
find the relevant solutions of Einstein equations.
General solutions of these equations are of course
not known, so a less ambitious program consists in trying to
find out a simple model for which we do have solutions! For this, we
will need to impose some constraints.

Before we even embark into describing the model itself, let us mention that
we need to confront the following limitations:

\noindent -- The Universe is unique by definition, so the usual methodology
of physics is not applicable as we can neither compare with other similar
objects to evaluate how generic our observations are nor redo experiments!

\noindent -- We are observing the Universe from a single location in both
space and time that we did not choose. In particular, this implies a question
about the history of the Universe and the specific moment we happen to
observe it. 

\begin{figure}[h]
\centering
\includegraphics[width=8.0cm,clip]{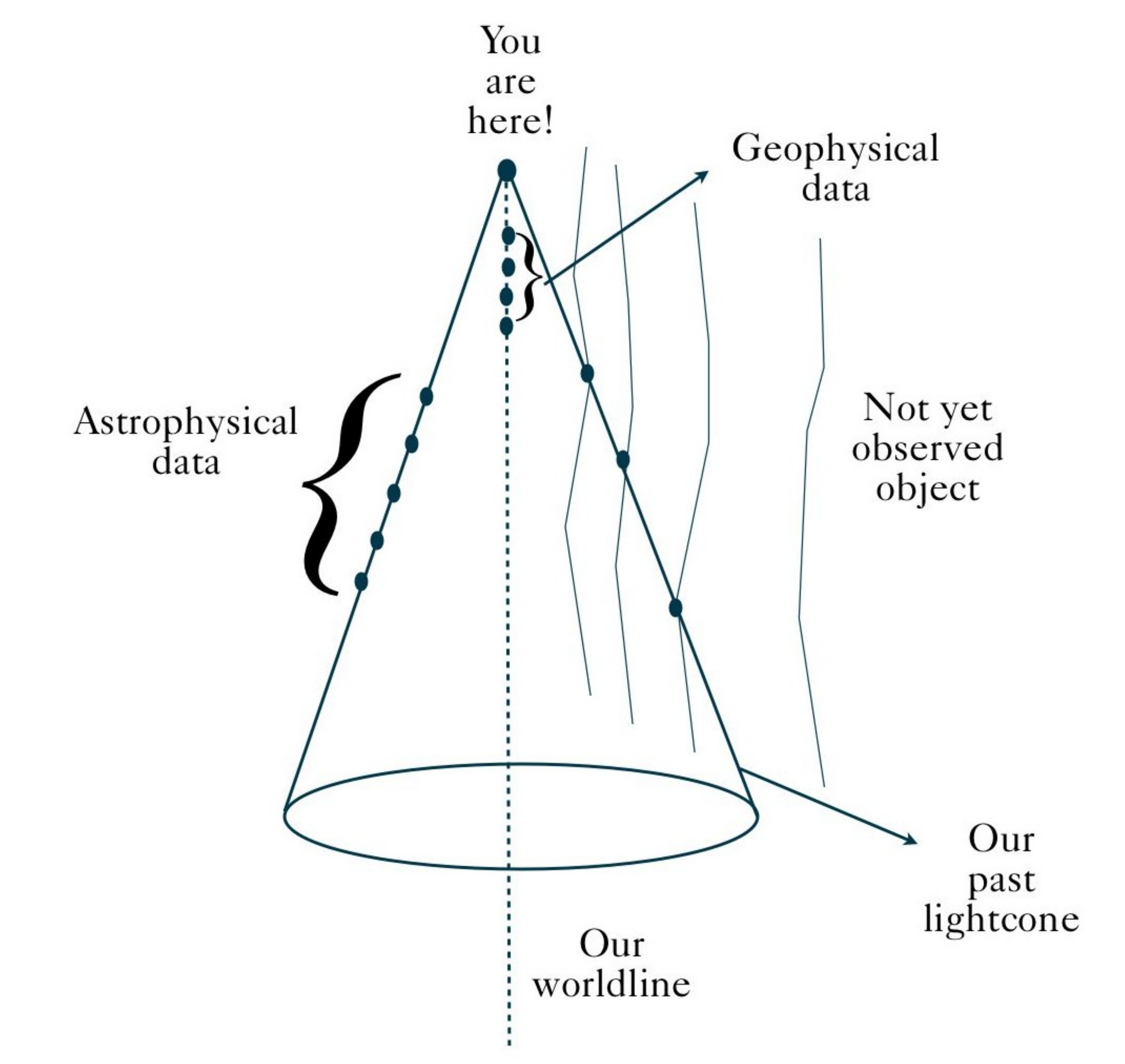}
\caption{Limitations due to our given and unchangeable position in
the Universe. We can only measure objects located inside our past
lightcone, would they be geophysical data on our worldline or astrophysical
data seen though its light emission, hence lying on the past lightcone.
Many objects, whose worldlines have not yet crossed our past lightcone,
are not visible yet, so a large fraction of the Universe is unreachable
to observation.}
\label{LocalObs}
\end{figure}

\noindent -- Observations,
as it turns out, are limited to our backward light cone, see figure~\ref{LocalObs}.

\noindent -- For a given set of data, there are possibly many space-times
corresponding to the observations. Again, as we have only one set of
observations
and since we cannot redo the experiment consisting in having the
Universe evolving from the Big-Bang to now, we have no way to make sure
our interpretation of the data is the correct one. We need to make some
hypothesis on the nature of the structure of space-time and verify those.
Only the large number of repetitions of observational data can reduce the
risk of confusion between different models. Nowadays, we have so many
data explained by one single model that it has become extremely difficult,
if not altogether impossible, to come up with a different but equally
successful model.
One can already notice, at this point, that a further complication with any
model of the Universe is that most of the hypothesis are hard to verify,
since we can actually model \emph{our} Universe, \ie the observable one.
The description of the actual Universe, which may even be infinite in size,
does not belong to the realm of physics as we will never have access to it.
Indeed, the Universe is probably much larger than the observable Universe
over which we can collect data.

Let us now turn to what such a model consists of.

\subsection{The FLRW model}

The now-standard model of cosmology is called after Alexander Friedmann,
Georges Lema\^\i{}tre, Howard Percy Robertson and Arthur Geoffrey Walker,
who first introduced and discussed the corresponding metric and applied
it to the Universe. The corresponding metric and space are called accordingly
the FLRW metric, although for some unclear reason the "L" is often
omitted...

\subsubsection{Theoretical hypothesis}

The FLRW model is mostly based on 4 basic assumptions:

\paragraph{Theoretical prejudice and framework}

Gravity is the leading force driving the dynamics, and we shall
describe it by means of the General Relativity (GR) theory of
gravitation. Based
on the equivalence principle, it is an
extremely well tested theory\footnote{GR is accurate at the level of
$10^{-12}$, \ie it compares with QED as far as predictions are concerned.
This constraint is obtained by measuring the orbital period variation of the
binary pulsar system which implies emission of gravitational waves in exact agreement with GR. This measurements led to its discoverers,
Russell A. Hulse and Joseph H. Taylor, sharing the Nobel prize for physics
in 1993.}, in particular in the Solar System in which it serves as a reference
for any alternative theory (scalar-tensor or MOND, for instance).

Gravity it is the only known unscreened long-range force, and thus appears
to be very well suited to describe the largest scales and even
the Universe as a whole. In assuming GR to hold on these
scales, we suppose the locally derived laws of physics apply and
can be extrapolated. On the other hand, if anything were to go
wrong in our description, that would lead to a natural testing ground
for GR.

The other interactions are
assumed to be well described by the standard
model of particle physics minimally coupled to gravity. This
is achieved through the
metric factor present in particular in the derivative terms: for
instance, for a
scalar field, the microscopic Lagrangian will contain a term of the
form $\mathcal{L}_\mathrm{kinetic}= -\frac12 g^{\mu\nu} \partial_\mu
\phi\partial_\nu\phi$. Hence, the fundamental action we shall be
interested in reads
\begin{equation}
\mathcal{S} = \int\dd^4x \sqrt{-g}\, \left\{ 
\frac{1}{16\pi\GN} \left(R-2\Lambda \right)
+ \mathcal{L}_\mathrm{matter}\left[ \phi(x),\psi(x),\cdots, 
g_{\mu\nu}\right]\right\}.
\label{ActionTot}
\end{equation}
It should be noticed again that any departure from this theoretical
framework translates into observations different from the expectation,
hence providing a way to test the validity of Eq.~(\ref{ActionTot}). In
particular, scalar-tensor theories that would be equivalent to GR on
Solar-System scales or for large cosmological times could originate
(either in scale or time) very far from GR, and that could lead to
observable consequences.

Reasons for doubting the validity of GR in astrophysics and cosmology
include the flat rotation curves of galaxies and the currently observed
acceleration of the Universe. At least at a phenomenological level however,
they can be described by GR providing extra "stuff" (dark matter and
energy) is added to the matter content to which I now turn.

\paragraph{Matter content}

Once the theoretical framework is fixed, one needs to set the matter content,
\ie the right hand side of Einstein equation (\ref{Einstein}). Observations,
made only over luminous matter (and hence not precluding a priori any
dark component) on the past light cone, reveal a single class of objects,
the luminous ones!
Therefore, we need to model not only those observed objects, but also any other
component that we would not actually be able to see.

The typical distance scales involved are the galaxy characteristic size, of
the order of $10^6$ light-years, and that of galaxy cluster, namely
$10^8$ light-years. Hence, we do expect some amount of clumsiness
on scales of these orders: the large scale structure of the Universe, being
supposedly insensitive to the small scale effects, will then be defined
on scales larger than $10^8$ light-years. On these scales, we will suppose
the matter content to form a perfect fluid with normalized 4-velocity $u^\mu$
($g_{\mu\nu} u^\mu u^\nu = -1$ with a metric with signature -2) and
stress energy tensor
\begin{equation}
T^{\mu\nu} = \left( \rho+p\right) u^\mu u^\nu + pg^{\mu\nu},
\label{Tmunu}
\end{equation}
where $\rho (\bm{x},t)$ and $p(\bm{x},t)$ are the energy density
and pressure. The dynamics is usually imposed by setting $\nabla_\mu
T^{\mu\nu}=0$, but this relates the time evolution of $\rho$ and $p$
in a contrived way: one needs to impose another relation, called the
equation of state, expressing the pressure as a unique function of the
energy density. In practice, a few simple cases are considered,
always assuming a linear relationship, i.e. $p=w\rho$, with $w$ a
constant called the equation of state parameter.

Observations reveal the typical relative velocity between
galaxies (the point particles
in the fluid element description) to be of the order $\langle v_\mathrm{gal}
\rangle \simeq 200\mathrm{km}\cdot\mathrm{s}^{-1}\sim 10^{-3}$ in units
of the speed of light. Therefore, the mean kinetic energy relative to
the mass can be evaluated as $E_\mathrm{kin}/\rho \simeq \frac12
\langle v^2_\mathrm{gal} \rangle \sim 10^{-6}$, and this also provides
a measure of the numerically expected value of the ratio between
pressure and density: $p/\rho \sim \frac13 \langle v^2_\mathrm{gal}
\rangle \sim 10^{-6}$. Therefore, the fluid made up with the galaxies
and any similar behaving fluid (dark matter) will be described by a
pressureless gas, i.e. $w_\mathrm{m} \sim 0$.

We also observed that the Universe is filled with some amount of
radiation, whose stress energy tensor is traceless, thus implying
$w_\mathrm{r}=\frac13$. Finally, a cosmological constant term
can be described by writing $T^{\mu\nu}_\Lambda = -(\Lambda/
8\pi\GN) g^{\mu\nu}$, and a direct comparison with (\ref{Tmunu})
then shows that this implies $p_\Lambda=-\rho_\Lambda$, in
other words $w_\Lambda=-1$. Amazingly, this extremely simple
set of 3 constant equation of state fluids suffices to describe the
evolution of the Universe for the previous 13.7 billions of years
with percent accuracy!

\paragraph{Symmetries}

\begin{figure}[t]
\centering
\includegraphics[width=12.0cm,clip]{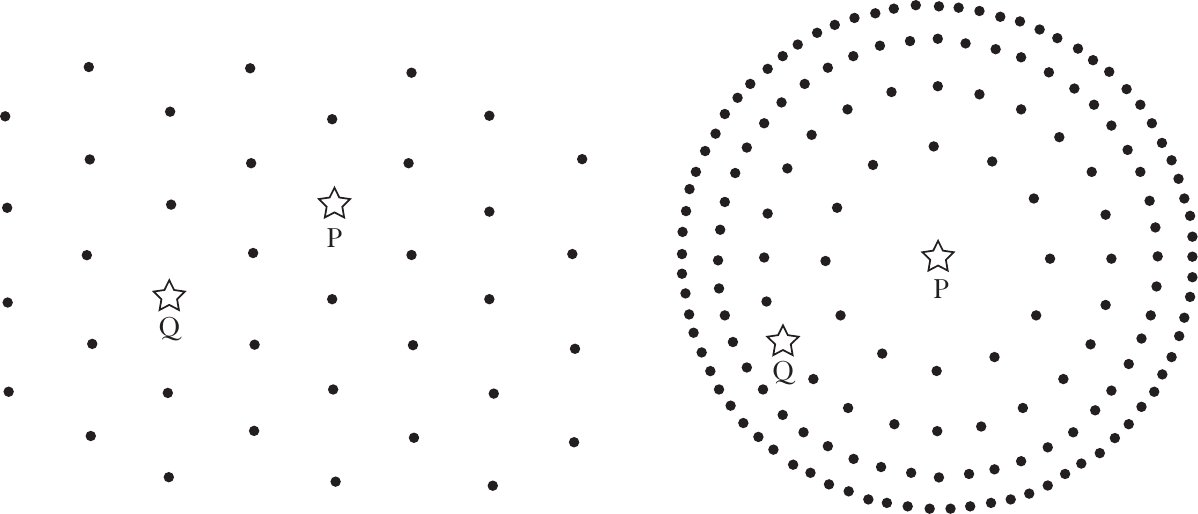}
\caption{We observe an isotropic distribution of matter, and that can
correspond either to an homogeneous distribution (left) or
to a spherical shell-like structure centered on
our location (right). Copernican principle
demands the first option, but such a philosophical posture,
however well justified, needs be verified experimentally; this can
nowadays be done by 3 dimensional observations using redshift
data.}
\label{Isotropic}
\end{figure}

Without symmetry assumptions, it is impossible to solve the full
GR equations, even with a given (and simple) stress energy tensor
such as that presented above, and so one needs to make even
more simplifying assumptions, again based on observations. Those
reveal the distribution of matter and radiation to be essentially
the same in all directions. In other words, we see a space which
appears isotropic. Figure \ref{Isotropic} then implies at least two
options following from these observations, of which the simplest
is homogeneity (but spherical symmetry has also been studied),
to which I will stick for now on. It should be emphasized that
both homogeneity and isotropy are concepts whose validity
in cosmology makes only statistical sense, and it is in this
sense that they must be verified whenever possible.

\begin{figure}[h]
\centering
\includegraphics[width=12.0cm,clip]{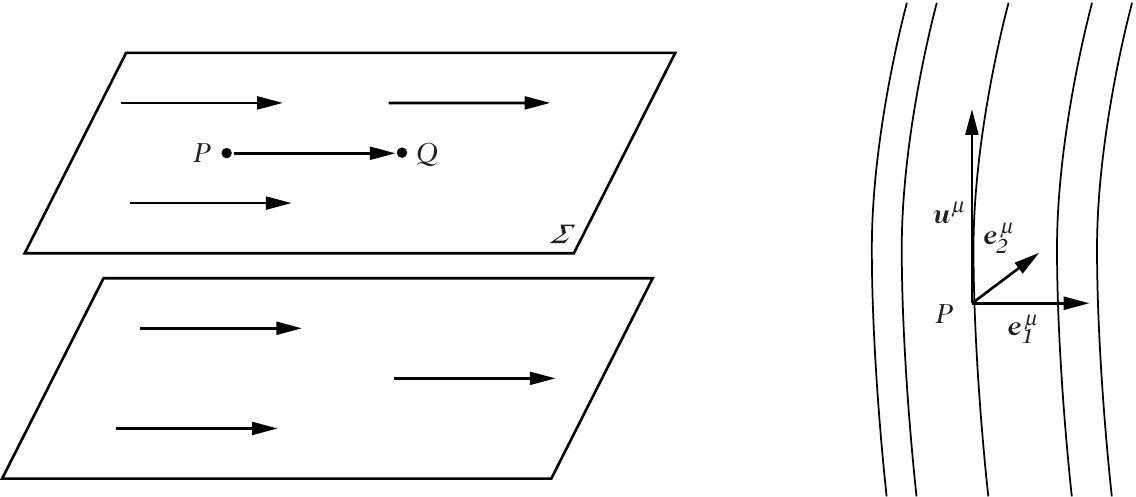}
\caption{Homogeneity (left) and isotropy (right) can be given exact
mathematical meaning in GR: any two points $P$ and $Q$ can be
related in an invariant way through an isometry in the hypersurfaces
$\Sigma_t$ (homogeneity $\to$ generalization of translations in space),
and for any point $P$ and two spacelike orthonormal vectors $e^\mu_1$
and $e^\mu_2$ such that $e^\mu u_\mu=0$, there exists an
isometry transforming $e_1 \leftrightarrow e_2$ (isotropy $\to$
generalization of rotations).}
\label{IsoHom}
\end{figure}

At the level of Newton classical theory, homogeneity, stating that
each point of space is similar to any other at each instant of time,
is well defined. In GR however, the previous sentence is absolutely
meaningless, and requires that a $3+1$ (space and time) slicing
is done, hence generating a one parameter (time $t$)
family of spacelike hypersurfaces $\Sigma_t$. Homogeneity
then is rephrased by saying that for any two points in $\Sigma_t$,
there exists an isometry taking one to the other. Isotropy on the
other hand states that at each spacetime event, an observer
moving with the cosmic fluid (comoving observer) cannot distinguish
one direction of space from another one. One sees that the two
notions are quite intricate, even though one describes a property
of spatial hypersurfaces, while the other involves time development;
this is due to the nature of our observations, always done along a light
cone, hence mixing space and time measurements. Figure \ref{IsoHom}
clarifies these statements.

Assuming homogeneity and isotropy means that $\bm{h}(t) \equiv
\bm{g}\big|_{P\in\Sigma_t}$, restriction of the full metric $\bm{g}$ to the
hypersurface $\Sigma_t$, only depends on time $t$, so that $\Sigma_t$
is a 3 dimensional homogeneous and isotropic space with induced
metric
$$
h_{\mu\nu} = g_{\mu\nu} + u_\mu u_\nu,
$$
as can be checked by direct calculation of a vector lying in $\Sigma_t$
or orthogonal to it.

Let $^{(3)}R_{\mu\nu\alpha\beta}$ be the Riemann tensor associated
with the metric $h_{\mu\nu}$ on $\Sigma_t$. By its definition (\ref{Riemann}),
it is symmetric in the exchange of the pairs of indices $\{\mu\nu\}$ and
$\{\alpha\beta\}$ and so can be seen as a map $L$ of the vector space
of 2-forms onto itself: setting $A\equiv \{\mu\nu\}$ and
$B\equiv\{\alpha\beta\}$, the matrix $L_A^{\ B}$ representing the
3 dimensional Riemann tensor is a symmetric matrix and thus
diagonalizable. If its eigenvalues were not all equal, then one of them could
be used to generate a privileged direction, in contradiction with the
hypothesis of isotropy. Hence, we have $L=K \setUni$, where
$K\in \setR$ can only depend on time, and $\setUni$ the identity in
the relevant space.

Moving back to 4 dimensional indices, we can write the 3 dimensional
Riemann tensor in the form
\begin{equation}
^{(3)}R_{\mu\nu\alpha\beta} = K (t) \left( h_{\alpha\mu} h_{\nu\beta}
-h_{\alpha\nu} h_{\mu\beta}\right).
\end{equation}
Let us see the meaning of this expansion for $K>0$ to begin with, and
embed the 3 dimensional space in a 4 dimensional Euclidian space with
coordinates $x$, $y$ $z$ and $w$. A
constant positive curvature space is a 3-sphere of radius $a$ whose
point locations are given by
\begin{equation}
x^2 + y^2 + z^2 + w^2 = a^2.
\label{xyzwR}
\end{equation}
In spherical coordinates defined by
\begin{equation}
\left\{ \begin{array}{rcl}
x&=&a \cos\chi,\\
y&=&a \sin\chi \cos\theta,\\
z&=&a \sin\chi\sin\theta\cos\varphi,\\
w&=&a \sin\chi \sin\theta \sin\varphi,
\end{array}
\right.
\end{equation}
differentiation of Eq. (\ref{xyzwR}) then provides the 3 dimensional metric
in the form
\begin{equation}
\dd^{(3)}\!\!s^2 \equiv 
\left( \dd x^2 + \dd y^2 + \dd z^2 + \dd w^2\right) \Big|_{a=\mathrm{cst}}
= a^2(t) \left[ \dd \chi^2 + \sin^2\chi \left( \dd \theta^2 + \sin^2\theta \dd
\varphi^2 \right) \right],
\label{inducedK1}
\end{equation}
where in the last line we have put back the possible time dependence of
the overall spatially constant curvature.

Similar considerations with negative (3-hyperboloid) or flat (Euclidian)
space permit to rewrite the overall 4 dimensional metric in the special
FLRW (at last!) form
\begin{equation}
\dd s^2 = g_{\mu\nu} \dd x^\mu \dd x^\nu = \left( h_{\mu\nu} -u_\mu
u_\nu\right) \dd x^\mu \dd x^\nu
=- \left( u_\mu \dd x^\mu\right )^2 + h_{\mu\nu} \dd x^\mu \dd x^\nu
=-\dd t^2 + a^2(t) \gamma_{ij} \dd x^i \dd x^j,
\end{equation}
where the spacelike part of the metric is
$$
\gamma_{ij}\dd x^i \dd x^j = \dd \chi^2 + f^2_\Ka \left(\chi\right)
\dd\Omega^2, \ \ \ \mathrm{with} \ \ \ f_\Ka = \Ka^{-1/2} \sin \left( \sqrt{\Ka}
\chi \right),
$$
where $\dd\Omega^2=\dd \theta^2 + \sin^2\theta \dd\varphi^2$ is the
usual solid angle element and the function $f_\Ka$ is to be continued for
vanishing ($\lim_{\Ka\to 0}f_\Ka = \chi$) or negative [$f_{\Ka<0} \to \left(
-\Ka\right)^{-1/2} \sinh \left( \sqrt{-\Ka} \chi \right)]$ values of $\Ka$.

In the above relations, we have written $\Ka$ to distinguish from the
function $K(t)$ giving the 3 dimensional Riemann tensor. In fact, it is
always possible to renormalize the spatial coordinates in such a way
that the scale factor $a(t)$ has the dimension of length, so that $\Ka$
can take one of the possible values $\Ka\in \{ 0,\pm 1\}$. This is the choice
we will assume for now on.

To finish this paragraph, I suggest to the reader to try and show, as
an exercise, that the spatial metric can be cast in the equivalent
forms
\begin{equation}
\gamma_{ij}\dd x^i \dd x^j = \frac{\dd r^2}{1-\Ka r^2} + r^2 \dd\Omega^2
= \frac{\dd \ell^2 + \ell^2 \dd\Omega^2}{\left(1+\displaystyle{\Ka}{4}
\ell^2 \right)^2}
\end{equation}
by means of changes of coordinates $\chi\to r \to \ell$ to be determined.

\paragraph{Topology}

GR is a local theory which thus says nothing about the global structure
of the Universe. String theory, of which GR is a low energy limit,
teaches us that some dimensions may be compact, and in fact need
be so in order for space to appear 3 dimensional on the scales available
to experiments. Therefore, in principle at least, it is possible that the
large dimensions we happen to live in could also be compact, leading
to a non trivial topology. Although this has been studied in details, I
shall not consider any further this hypothesis for at least two reasons,
one experimental and the other theoretical.

First, there is no data, to date, that would induce us to think a large
scale non trivial topology is needed. Of course, some compact models
can be made compatible with the data, or even improve the fit, but
they are degenerate with other models and the improvement is not
really statistically significant.

Second, an argument in favor of compact large dimensions could
be to invoke compactness for {\sl all} dimensions; in this case, the
expected phase of inflation would make the large dimensions much
larger than the current Hubble scale (size of the observable universe),
unless a disturbingly severe fine-tuning is applied. If a non inflationary scenario
is implemented, then another fine-tuning is necessary in order to
explain why the lattice size of the compact dimension should be,
today, of the order  of the Hubble scale (only case not yet ruled out
by the data but still leading to observable predictions).

\vskip5mm

Having settled the framework, let me move on to the dynamics of
our Universe.

\subsection{The dynamical Universe}

The framework developed above permits to write down explicitly
the Einstein equations as a set of relations between a very small
subset of dynamical quantities, namely the scale factor $a(t)$
and the density of the fluid $\rho(t)$. In order to derive these
equations, we need to calculate all the relevant geometrical quantities.

\subsubsection{Geometrical quantities}

The Einstein equations involve in a non trivial way the Riemann tensor
and its byproducts, namely the Ricci tensor and scalar and the Einstein
tensor itself. Those are all built from the metric connections and
ultimately from the metric itself. It turns out that the cosmic time $t$
introduced earlier is not the most convenient time parameter, especially
when the spatial sections are flat (which is observationally the case), and
we usually introduce a dimensionless time, called the {\sl conformal time}
$\eta$ as it renders the metric conformally flat. It is related to the
cosmic time by 
\begin{equation}
a \dd \eta = \dd t \ \ \ \ \Longrightarrow \ \ \ \dd s^2 = a^2(\eta) \left( -\dd \eta^2
+ \gamma_{ij} \dd x^i \dd x^j\right) \underset{\Ka\to0}{=} a^2(\eta) \left( 
-\dd\eta^2 + \dd x^2 + \dd y^2 + \dd z^2 \right),
\label{ConfTime}
\end{equation}
where in the last stage we have taken the limit $\Ka\to0$ to
make the Minkowski metric apparent.

To simplify matters, we define derivatives with respect to times as
$\dot f\equiv \dd f/\dd t = \dd f/(a \dd \eta) \equiv a^{-1} f'$. Then, setting
$H\equiv \dot a/a$ and $\Hu=a'/a = \dot a = a H$,
we obtain the only non vanishing
connection coefficients as
\begin{equation}
\Gamma^t_{ij} = a^2 H \gamma_{ij}, \ \ \ \ \Gamma^i_{tj} = H\delta^i_j \ \ \ 
\mathrm{and} \ \ \ \Gamma^i_{jk} = \gamma^i_{jk}
\label{Chrt}
\end{equation}
in cosmic time, and
\begin{equation}
\Gamma^\eta_{\eta\eta} = \Hu, \ \ \ \ \Gamma^\eta_{ij} = \Hu\gamma_{ij} \ \ \ 
\mathrm{and} \ \ \ \Gamma^i_{\eta j} = \Hu \delta^i_j,
\label{Chreta}
\end{equation}
in conformal time. From these, one derives the non vanishing Ricci tensor
components
\begin{equation}
R_{tt} = -3\frac{\ddot a}{a},  \ \ \ \ \ R_{ti} = 0 \ \ \ \ \mathrm{and}
\ \ \ \ \ R_{ij} = a^2 \gamma_{ij} \left( \frac{\ddot a}{a} + 2H^2 + 2 \frac{\Ka}{a^2}
\right),
\end{equation}
leading to the Ricci scalar $R=6\left( H^2 + \ddot a/a + \Ka/a^2 \right)$.

Combining these, we finally obtain the Einstein tensor as
\begin{equation}
G^t_t = -3\left( H^2 + \frac{\Ka}{a^2}\right), \ \ \ \ G^t_i = 0 \ \ \ \ \ \mathrm{and}
\ \ \ \ \ G^i_j=-\delta^i_j \left( 2\frac{\ddot a}{a} + H^2 + \frac{\Ka}{a^2}
\right).
\label{EinsteinFL}
\end{equation}
This provides the left hand side of Einstein equations (\ref{Einstein}).

\subsubsection{Friedmann equations and the cosmological parameters}

With the geometric quantities derived for the FLRW metric, and the stress
energy tensor (\ref{Tmunu}) restated in matrix form as $T^{\mu\nu} =
\mathrm{diag}\left( -\rho, p,p,p\right)$, it now remains to equal it to
(\ref{EinsteinFL}) to obtain the Friedmann equation, which the
reader will straightforwardly check they can be cast in the form
\begin{equation}
H^2 = \frac{8\pi\GN}{3} \rho - \frac{\Ka}{a^2} + \frac{\Lambda}{3}
\label{H2t}
\end{equation}
for the constraint, and
\begin{equation}
\frac{\ddot a}{a} = - \frac{4\pi\GN}{3} \left( \rho+3p\right)
+ \frac{\Lambda}{3}
\label{att}
\end{equation}
for the dynamics. Deriving Eq. (\ref{H2t}) with respect to time, taking
into account Eq. (\ref{att}) and reshuffling the various terms involved
yields the fluid conservation equation
\begin{equation}
\dot \rho + 3 H \left( \rho+p\right) = 0 \ \ \ \ \ 
\Longleftarrow \ \ \ \ \ \nabla_\mu
T^{\mu\nu} =0,
\label{dTt}
\end{equation}
as expected since the latter conservation is not independent of the Einstein
equations from which (\ref{H2t}) and (\ref{att}) stem.

In terms of conformal time, the previous set of equations read
\begin{equation}
\rho' + 3\Hu \left( \rho+p\right) =0,
\label{dTeta}
\end{equation}
for the conservation equation, 
\begin{equation}
\Hu^2 + \Ka = \left(\frac{8\pi\GN}{3} \rho +\frac{\Lambda}{3}\right)a^2
\label{calH2}
\end{equation}
for the constraint, and finally
\begin{equation}
\Hu' = \left[ - \frac{4\pi\GN}{3} \left( \rho+3p\right)
+ \frac{\Lambda}{3} \right] a^2.
\label{Hp}
\end{equation}

There exists a special solution, which happens to be realized in our
Universe, at least so seem to say the data, namely that for which the
spatial curvature $\Ka$ vanishes. It defines a density, called the critical
density $\rho_\mathrm{c}$ given by
\begin{equation}
\rho_\mathrm{c} \equiv \frac{3H^2}{8\pi\GN} \ \ \ \ \Longrightarrow\ \ \ \ 
\Omega \equiv \frac{\rho}{\rho_\mathrm{c}},
\label{rhoc}
\end{equation}
in terms of which one can express all densities in a dimensionless way. For
each fluid component but the cosmological constant, one can set
$\Omega_a = 8\pi\GN\rho_a/(3H^2) = \rho_a/\rho_\mathrm{c}$; we
also introduce an equivalent curvature "density" as $\Omega_\Ka =- \Ka
/(a^2 H^2)$ and finally $\Omega_\Lambda = \Lambda/(3H^2)$, and then
the Friedmann constraint simply reads:
\begin{equation}
\sum_a \Omega_a + \Omega_\Lambda + \Omega_\Ka =1,
\label{FriedOm}
\end{equation}
so the Friedmann equation is understandable as an energy budget:
all possible contributions basically sum up to 100\%!
Numerically, the Hubble constant today is measured to be of the order
of $H_0 = 100 h \,\mathrm{km}\cdot\mathrm{s}^{-1} 
\cdot\mathrm{Mpc}^{-1}$, where $h= 0.704 \pm 0.025$. Similarly,
the relative densities are also measured in units of the critical
density, estimated as $\rho_\mathrm{c} \simeq 1.9\times 10^{-29}
\mathrm{g}\cdot \mathrm{cm}^{-3}$; they frequently are found
expressed as $\rho^0_{i} = \Omega^0_{i}h^2$ to account for the
indeterminacy of the Hubble expansion rate as well as on the
density parameter itself, the subscript ``$0$'' meaning the present-day
value.

\subsection{Special solution: matter and radiation}

\begin{figure}[h]
\centering
\includegraphics[width=8.0cm,clip]{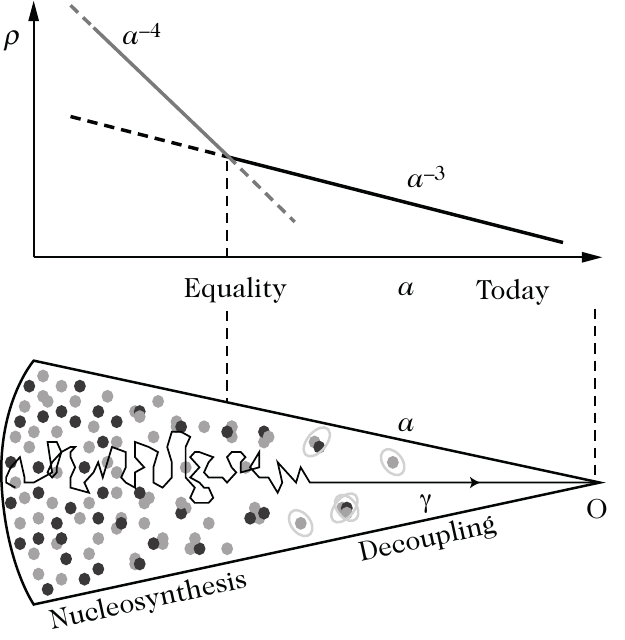}
\caption{Top -- Evolution of densities: the Universe begins dominated
by radiation, whose density decreases faster than that of
matter, so the latter ultimately
dominates. Not shown is the final phase of domination by
a cosmological constant which, as its name indicates, behaves as a
constant. The point at which radiation and matter contribute equally is,
not surprisingly, called {\sl equality}. Bottom -- On the same scale,
matter density is depicted together with a typical light ray, whose mean
free path is initially much shorter than the Hubble scale, as e.g.
during nucleosynthesis; as the matter density gets smaller and smaller,
the mean free path eventually becomes larger than the Hubble scale after
what is therefore denoted {\sl decoupling}. The Universe becomes transparent
to this radiation we now observe as the microwave background.}
\label{rhos}
\end{figure}

With a varying equation of state $w(t)$ and a scale factor $a(t)$,
which is a monotonic function of time, it is always possible to
parameterize all functions of time as functions of $a$, and in
particular $w$. On can then formally integrate the conservation
equation as
\begin{equation}
\rho[a(t)] = \rho_\mathrm{ini} \exp \left\{ -3 \int \left[ 1+w(a)\right]
\dd \ln a\right\} \underset{w\to\mathrm{cst}}{=} \rho_\mathrm{ini}
\left( \frac{a}{a_\mathrm{ini}}\right)^{-3(1+w)},
\label{consInt}
\end{equation}
which gives an exact solution for the constant equation of state
situation. This is precisely the case when matter ($w\to w_\mathrm{m}=0$) or
radiation ($w\to w_\mathrm{r}=\frac13$) dominates over everything else. Eq. (\ref{consInt})
then shows that matter scales as $\rho_\mathrm{m} \propto
a^{-3}$, as expected from mass
conservation in an expanding volume, while radiation gets an extra
power, scaling as $\rho_\mathrm{r} \propto a^{-4}$, due to the 
redshift of its wavelength. Now consider an initial condition consisting
of given relative amounts of matter and radiation. When the Universe
begins its evolution, with a small value of the scale factor, radiation
dominates and the total density is $\rho_\mathrm{tot} \sim \rho_\mathrm{r}$
until it gets caught up by the dustlike matter. This remarkably accurate
picture for the Universe density evolution is illustrated in figure \ref{rhos}.

The meaning of the equation of state is clarified when one considers
a perturbation propagating in the fluid. As is well known in fluid 
dynamics and as we shall also discuss later, the sound velocity
$c_\mathrm{s}$ is given\footnote{In fact, it should be partial
derivative for constant entropy.} by $c_\mathrm{s}^2 = \dd p/\dd \rho = p'
/\rho'$. It can be shown (and the reader is encouraged to do so!),
that the relation
\begin{equation}
w' = -3\Hu \left(1+w\right) \left( c_\mathrm{s}^2 - w\right)
\label{wp}
\end{equation}
holds, so that a constant equation of state means $w=c_\mathrm{s}^2$.

With the solution for the density as a function of the scale factor and
the equation of state given, it is an easy matter to solve the
Friedman equation. For a vanishing spatial curvature $\Ka=0$, one finds
that if $w\not= -1$, the solution goes like 
\begin{equation}
a\propto t^{2/[3(1+w)]} \propto \eta^{2/(1+3w)} \ \ \ \ \Longrightarrow \ \ \ 
a_\mathrm{r} \propto \sqrt{t} \propto \eta \ \ \ \hbox{and} \ \ \ \
a_\mathrm{m} \propto t^{2/3} \propto \eta^2,
\label{aw}
\end{equation}
where I emphasized the particular pressureless dust and radiation
dominated solutions. In the special case of a cosmological constant
with $w=-1$, one finds
\begin{equation}
\dot\rho=0 \ \ \ \Longrightarrow \ \ \ H=\hbox{cst} \ \ \ \Longrightarrow
\ \ \ a\propto \ex^{Ht} \propto \frac{-1}{H\eta},
\label{adS}
\end{equation}
and one has an exponentially accelerated expansion; note in that case,
which will later correspond to the inflationary situation, that the conformal
time is negative, with the end of inflation being for the limit when
$\eta\to 0$.

\subsection{Limitations of the standard model}

The model developed above gives a quite accurate description
of the history of the Universe, but its success actually raises a
few questions that find no answer in its own framework.

\subsubsection{Puzzles}

\paragraph{Singularity}

The first troubling issue is also the only one that has, in the
inflationary paradigm, not received any answer, namely the
fact that whatever solution of Einstein equations one comes
up with that fits the available observational data does begin
with a primordial singularity: at some point in the past, there
is always a time $t_\mathrm{sing}$
at which $a(t_\mathrm{sing})\to 0$, meaning
all the geometrical tensors diverge, so the theory itself simply
does not make sense anymore! One can however argue that
GR is not designed to handle extremely high energies so that
a cutoff, at the string or Planck scale, should be applied, above
which the theory will (wishful thinking) be regular.

\paragraph{Horizon}

\begin{figure}[h]
\centering
\includegraphics[width=12.0cm,clip]{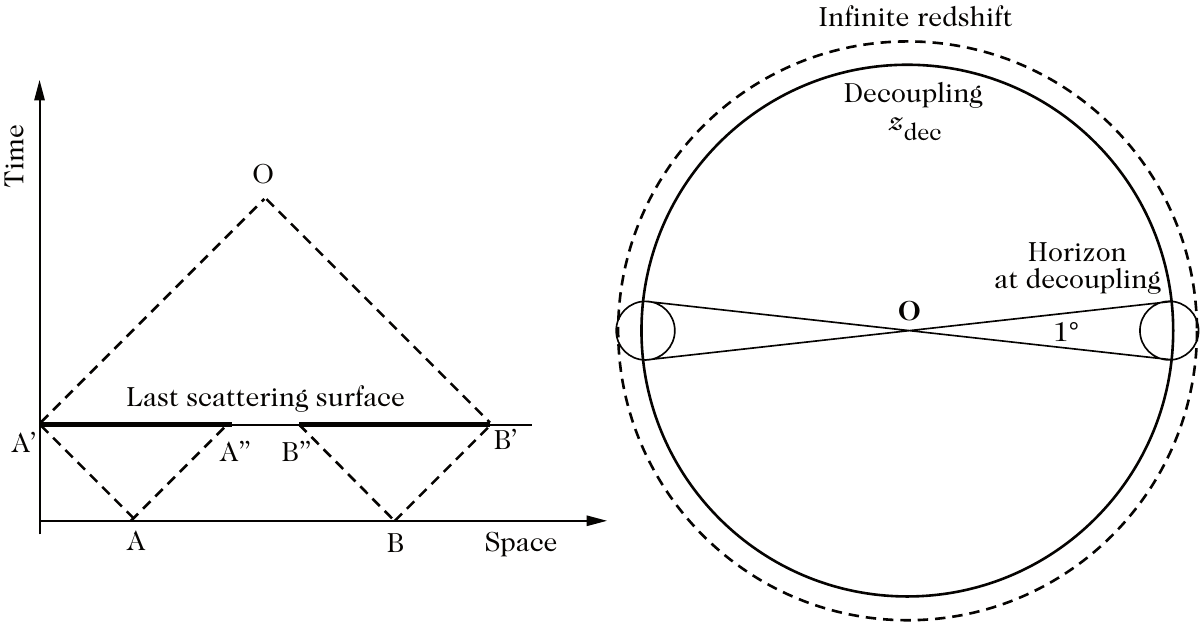}
\caption{The horizon problem. Left -- spacetime diagram beginning
at the Big-Bang (i.e. the singularity). Light emitted in $A$ and $B$
in all possible directions reach regions at the last scattering surface
(surface of decoupling when light stopped scattering and started
propagating unaffected) that never were in causal contact. Yet, 
they appear to have exactly the same physical properties. Right --
Angular representation of the same thing: the Big-Bang singularity is
now represented from our point of view by the infinite redshift
sphere. Calculating the horizon size at decoupling gives one degree
on the sky, which is thus the maximal angular scale over which one
might expect to measure an isotropic distribution.}
\label{Horizon}
\end{figure}

The question of the horizon is more involved in a way, as no
hand-waving argument can be similarly invoked to cure it. It
relies on the observed fact that light emitted at decoupling (see
figure \ref{rhos}) is homogeneous up to $10^{-5}$. Although
this looks like a mere consequence of the cosmological principle,
it is actually weird because of the previously discussed
singularity problem: the existence of a primordial singularity
implies a Big-Bang, i.e. a point in time at which the Universe
expansion starts, so that there was a finite amount of time for
a priori initially causally disconnected regions to thermalize. When
one estimates the number of such regions, one finds some $10^{5}$
of those at decoupling, implying a predicted isotropy over angular
scales smaller than roughly one degree on the sky only! Figure
\ref{Horizon} illustrates the issue.

\paragraph{Flatness}

Finally, the flatness problem is based on the fact that the observed
flat spatial section ($\sum_a\Omega_a+\Omega_\Lambda =1$, i.e.
$\Omega_\Ka=0$) is actually an unstable fixed point: in the absence
of a dominating cosmological constant, deriving
Eq. (\ref{FriedOm}) with respect to the scale factor yields
\begin{equation}
\frac{\dd\Omega_\Ka}{\dd\ln a} = \left( 3w+1\right) \left( 1-\Omega_\Ka
\right) \Omega_\Ka,
\label{dOmda}
\end{equation}
whose solution, for a constant equation of state, reads
\begin{equation}
\Omega_\Ka \left(a_\mathrm{obs}\right) =
\Omega_\Ka^\mathrm{ini} \left[ \left( 1- \Omega_\Ka^\mathrm{ini}\right)
\left(\frac{a_\mathrm{obs}}{a_\mathrm{ini}}\right)^{1+3w}
+ \Omega_\Ka^\mathrm{ini} \right]^{-1},
\label{solOm}
\end{equation}
where $\Omega_\Ka^\mathrm{ini} =\Omega_\Ka (a_\mathrm{ini})$. In
order to observe now $\Omega_\Ka \left(a_\mathrm{obs}\right) \lta 0.1$,
one then needs to demand that at equality ($a_\mathrm{obs}/
a_\mathrm{eq}\simeq 10^4$), $|\Omega_\Ka^\mathrm{eq}|\lta 3\times
10^{-5}$. which represents already a quite substantial amount of
fine tuning if $\Omega_\Ka$ is to be an arbitrary initial condition. It
becomes even worse of course if one assumes initial conditions
ought to be imposed at one Planck time after
the Big-Bang singularity itself, as the
requirement then becomes $|\Omega_\Ka^\mathrm{Planck}| \lta
10^{-60}$: this is an unacceptably large amount of fine tuning!

\subsubsection{Categories of solutions}

There are nowadays two categories of solutions as far as I can tell,
one widely accepted and usually set as part of the standard cosmological
paradigm, namely inflation, and a contender based on a contracting
phase and a bounce. Inflation does not address the singularity question, while
a bounce is in danger of producing too much shear during the
contraction. Moreover, inflation can be easily implemented using
a simple scalar field, a de Sitter phase actually being an attractor
in the equations of motion, while a bounce is almost impossible to
implement in the framework of 4 dimensional GR unless the spatial
curvature is positive. This makes inflation more appealing to most
people.

\paragraph{Inflation}

The Flatness problem originates from Eq. (\ref{dOmda}) and the fact
that $\Omega_\Ka=0$ is an unstable fixed point for this equation. In
fact, this is only true provided $w\geq -\frac13$. If this condition is not
fulfilled, as for instance is the case of a cosmological constant domination,
then $\Omega_\Ka=0$ becomes instead an attractor. So it suffices
to include a sufficiently long phase during which $\Omega_\Ka\to 0$,
then followed by the usual radiation and matter domination, to keep
$\Omega_\Ka$ close to zero even after a long time of regular
expansion. What is the meaning of this solution?

Equation (\ref{att}) in the absence of a cosmological constant shows that
if $w< -\frac13$, i.e. if $p<-\frac13 \rho$, then $\ddot a$ changes sign
and the expansion is accelerated. This is why this solution was called
{\sl inflation}. Very often, it is implemented by means of a "slowly
rolling" scalar field, dubbed {\sl inflaton},
i.e. a scalar field whose dynamics is dominated
by the potential term, naturally leading to $w\simeq -1$. As a result,
inflation is achieved by an almost exponential growth of the
scale factor.

Having a phase of accelerated expansion actually also solves without
any further assumption the horizon problem. Indeed, the horizon size
is a global quantity whose definition involves the overall history of the
Universe through 
\begin{equation}
d_\mathrm{H} = a(t) \int_{t_\mathrm{ini}}^t a^{-1}(T)\dd T,
\label{hor}
\end{equation}
where $t_\mathrm{ini}$ is the origin of times. Note that for
a power-law expansion such as during most of the history
of the Universe [see Eq. (\ref{aw})], i.e. if $a\propto t^\alpha$,
the horizon scales as $d_\mathrm{H} = t/(\alpha+1)$, which
is then roughly the same as the Hubble expansion rate $H^{-1}
=a/\dot a = t/\alpha$. This, plus the fact that "Hubble" and
"Horizon" begin with the same letter, has led to a confusion
in many works between the two quantities. I will come back to
that point later.

When an almost exponential phase of inflation takes place, the Hubble
radius is roughly constant, while the scale factor grows exponentially.
The horizon size is then
\begin{equation}
d_\mathrm{H} = \frac{1}{H}\left[ \ex^{H\left( t-t_\mathrm{ini} \right)}
-1\right] \gg H^{-1},
\label{Hexp}
\end{equation}
where the last inequality assumes $t\gg t_\mathrm{ini}$. What
happens then is that the horizon size grows much faster than the
Hubble scale so that all scales end up having time to be in
causal contact.

To be complete with the inflationary scenario, apart from its
prediction of an almost scale-invariant spectrum of primordial
fluctuations, I should like to mention that it is also the only known
way of naturally reducing any initial amount of anisotropy. However,
in order to reach an FLRW Universe, one also needs to impose
a sufficiently smooth initial patch, i.e. even though inflation substantially
alleviates the question of initial inhomogeneities, it
does not actually answer it.

\paragraph{Bouncing scenarios}

Eq. (\ref{dOmda}) can also be rewritten as 
\begin{equation}
\frac{\dd\Omega_\Ka}{\dd t} = - 2 \frac{\ddot a}{\dot a^3},
\label{Omt}
\end{equation}
emphasizing once again that a phase of accelerated ($\ddot a >0$)
expansion ($\dot a >0$) will drive $\Omega_\Ka$ to vanishingly
small values. Another way is possible, consisting in reverting all the
signs of the previous argument, therefore using a decelerated
($\ddot a <0$) phase of contraction ($\dot a <0$)! Since we observe
the current phase to be expanding, this implies that a transition between $H<0$
and $H>0$ took place, a bounce.

One might however argue that, as I said earlier, a positive spatial
curvature is required to implement such a bounce in GR. One quick
answer to this argument is that GR may not be valid at that time...
after all, inflation also requires some extension of GR to account
for the primordial singularity. Note in passing that the singularity
is easily gotten rid of in the bouncing scenario since time can be
pushed back as far as one wants, in principle even to infinity. This
actually also solve the horizon problem, since it can easily be
made infinite. To anyone not willing to extend GR in any way,
one could also argue that it suffices to have a very long contraction
phase during which $\Omega_\Ka\to 0$, and manage that the
bounce itself is not very asymmetric, so that even though
$\Omega_\Ka$ can grow large during the bounce, it will recover
a value after the bounce that is not very different from the one
it had before.

The bouncing scenario however is plagued with an anisotropy problem:
if one considers a initial shear, however tiny, it will grow very large
during either the contracting phase or during the bounce itself. This
is at least true in simple models, but more sophisticated scenarios
have been proposed that tame these unwanted large growth.

\paragraph{Structure formation: perturbation theory}

To obtain a complete description of the Universe, one would, at this point,
need to include thermodynamical evolution of all the relevant quantities,
taking into account interactions to describe in a reasonable way the phases
of nucleosynthesis for instance. I shall not embark in this direction, and will
instead concentrate on the perturbations over this background: those
in fact provide a bonus for the inflationary scenario, as by demanding
the inflaton to be in vacuum and allowing it to have quantum fluctuations,
the ensuing evolution transforms the Universe in a particle producer,
those particles then later behaving as large scale fluctuations seeding
the formation of structures.

\section{Perturbation theory: SVT and the gauge issue}

It is widely believed that large scale structures formed out of primordial
seeds upon which gravitational collapse acted to produce dense objects.
This theory is quite well verified, as numerical simulations starting with
an initial over-density function satisfying scale-invariant statistical
properties manage to reproduce the statistical properties observed
in large scale structure surveys. It lacks however a crucial ingredient:
what is the seed origin?

\subsection{Introductory remarks: the Jeans length\\ and Newtonian perturbation theory}

Newtonian physics allows to understand the origin of gravitational
collapse in the expanding Universe in a phenomenological way: one
simply assumes that Newtonian gravity holds, but also that the
Universe is expanding, i.e. that the actual distance $\bm{r}$ between
objects increases with time. One then has $\bm{r} = a(t) \bm{x}$, where
$\bm{x}$ is the relative position of the object in a local coordinate system
and $a(t)$ the scale factor discussed in the previous section. The total
velocity then consists in two pieces,
\begin{equation}
\label{v2}
\frac{\partial \bm{r}}{\partial t} \equiv \bm{v} = \dot a\bm{x} +
a \frac{\partial\bm{x}}{\partial t} =
\underbrace{a H \bm{x}}_{\rm background} + 
\underbrace{\bm{u}}_{\rm peculiar},
\end{equation}
where the first term represents the background cosmic flow involving
the Hubble rate $H$, and the second the peculiar velocity, i.e. a relative
velocity that one can treat as a perturbation.

Similarly, the density field is expanded as
\begin{equation}
\label{rho}
\rho\(\bm{x},t\) = \bar\rho\(t\)\[ 1+\delta\(\bm{x},t\)\],
\end{equation}
and the continuity equation becomes
\begin{equation}
\label{continuity}
\(\frac{\partial\rho}{\partial t}\)_r + \bm{\nabla}_r \cdot \( \rho\bm{v} \) =0 \ \ \ \ \Longrightarrow
\ \ \ \ \
\(\frac{\partial\rho}{\partial t}\)_x+3H\rho+\frac{1}{a}\bm{\nabla}_x\(\rho\bm{u}\)=0,
\end{equation}
where the `$3H$' term comes from changing the coordinate $\bm{r}$ to $\bm{x}$. To zeroth
order, Eq. (\ref{continuity}) implies $\dot{\bar{\rho}} + 3H\rho=0$, which merely reflects that
matter scales as $\bar\rho\propto a^{-3}$, while the first order yields
\begin{equation}
\label{deltadot}
\dot\delta +\frac{1}{a}\bm{\nabla}\cdot \[ \(1+\delta\) \bm{u}\] =0,
\end{equation}
where for now on we assume all spacelike derivatives are with respect to the
`comoving' coordinates $\bm{x}$; I shall accordingly subsequently omit the index $x$.

Combining Eq.~(\ref{deltadot}) with
\begin{equation}
\label{Euler}
\frac{\partial\bm{u}}{\partial t} + H\bm{u}+\frac{1}{a} \( \bm{u}
\cdot \bm{\nabla}\) \bm{u}+\frac{1}{a}\(\frac{1}{\rho}\bm{\nabla}P+\bm{\nabla}\Phi\)=0,
\end{equation}
which is nothing but the Euler equation for a fluid with pressure $P$ in a gravitational
potential $\Phi$ (satisfying the Poisson equation $\Delta\Phi = 4 \pi \GN \bar\rho\delta$)
in comoving coordinates, and linearizing, one finds
\begin{equation}
\label{pertGen}
\ddot \delta +\underbrace{2H\dot \delta}_{\rm expansion} -
\underbrace{\frac{\cs^2}{a^2} \Delta \delta}_{\rm pressure}
=\underbrace{4\pi\GN\bar\rho\delta}_{\rm gravity},
\end{equation}
showing the Newtonian evolution involves three distinct effects, namely
the damping of any perturbation due to the expansion, the propagation
of sound waves due to the pressure terms, and finally gravity itself. In Eq.~(\ref{pertGen}),
the sound velocity $\cs$ is defined as before
through $\cs^2\equiv\(\partial p/\partial\rho\)_S$
where the fluid entropy $S$ is held constant.

Expanding in Fourier modes ($\Delta \to -k^2$) and defining the physical
wavenumber $k_\mathrm{p} \equiv k/a$,  one obtains, forgetting for the
moment the expansion (i.e. setting $a\to\cte$)
\begin{equation}
\label{Jeans1}
\ddot \delta +\( \cs^2 k_\mathrm{p}^2 -4\pi \GN\bar\rho\)\delta = 0 \ \ \ \ \Longrightarrow\ \ \ \ 
\delta\propto\exp\[\sqrt{4\pi\GN\bar\rho\(1-\frac{\lambda^2_{_\mathrm{J}}}{\lambda^2}\)}t\]
\end{equation}
where wavelengths are defined by $\lambda=2\pi/k$, and $\lambda_{_\mathrm{J}}\equiv
\cs\sqrt{\pi/(\GN\bar\rho)}$ is the celebrated Jean's length separating regimes of
wave propagation and gravitational instability: for long wavelength, $\lambda >
\lambda_{_\mathrm{J}}$, the density is growing exponentially with time, signaling
a collapse, while for small wavelengths $\lambda < \lambda_{_\mathrm{J}}$, the
density oscillates as the sound wave propagates smoothly. Taking into account
the overall expansion does not change this result qualitatively, it merely changes
the functional dependence of the density with time, not the fact that there is a regime
of unlimited growth and another of oscillations.

Having settled the stage, let me now move to the real issue, namely that of
GR perturbations in FLRW Universe.

\subsection{3+1 decomposition}

For now on, I will mostly consider the conformal time $\eta$, in terms
of which the subsequent exposition is probably clearer. Then, the Friedmann
(Einstein) equations are given by (\ref{calH2}) and (\ref{Hp}). Having
completely fixed the background, we can now move on and expand
around this background.

\subsubsection{Perturbative expansion}

Our starting point is the action (\ref{ActionTot}) or, in practice, Einstein
equations (\ref{Einstein}). As we did obtain the homogeneous and isotropic
solution, we write it as $g^{(0)}_{\mu\nu}\( \eta\)$, leading to the
corresponding Einstein tensor $G^{(0)}_{\mu\nu}\( \eta\)$, itself
sourced by the background stress-energy tensor $T^{(0)}_{\mu\nu}\( \eta\)$.
We then write the full metric as
\begin{equation}
\label{FullMetric}
g^\mathrm{full}_{\mu\nu}\( \eta,\bm{x}\) = g^{(0)}_{\mu\nu}\( \eta\)
+\varepsilon g^{(1)}_{\mu\nu}\( \eta,\bm{x}\) 
+\frac12 \varepsilon^2 g^{(2)}_{\mu\nu}\( \eta,\bm{x}\)
+ \cdots,
\end{equation}
where the dots contain all higher order terms. We then assume that $\varepsilon$
is a small parameter, as data indicate it to be the case on sufficiently large
scales, i.e. on scales larger than roughly 200~Mpc. With Eq. (\ref{FullMetric}) and
the definition of the Einstein tensor, one can express it in the same way, namely
\begin{equation}
\label{EinsteinPerturbed}
G^\mathrm{full}_{\mu\nu}\( \eta,\bm{x}\) = G^{(0)}_{\mu\nu}\( \eta\)
+\varepsilon G^{(1)}_{\mu\nu}\( \eta,\bm{x}\) 
+\frac12 \varepsilon^2 G^{(2)}_{\mu\nu}\( \eta,\bm{x}\)
+ \cdots,
\end{equation}
where $G^{(0)}_{\mu\nu}\( \eta\)$ is given by (\ref{EinsteinFL}).

Similarly, we expand the stress-energy tensor (\ref{Tmunu}) as
\begin{equation}
\label{Ttotmunu}
T^\mathrm{full}_{\mu\nu}\( \eta,\bm{x}\) = T^{(0)}_{\mu\nu}\( \eta\)
+\varepsilon T^{(1)}_{\mu\nu}\( \eta,\bm{x}\) 
+\frac12 \varepsilon^2 T^{(2)}_{\mu\nu}\( \eta,\bm{x}\)
+ \cdots,
\end{equation}
which amounts to expanding $\rho$, $p$ and the fluid vector $u_\mu$.
Providing the series in powers of $\varepsilon$ makes sense, it now
suffices to expand both sides of Einstein equations and identify the
terms, order by order.

In practice, there is no $\varepsilon$ parameter, and we merely expand
all relevant quantities as ``background'' + ``something small'' which we
then calculate. The metric $g_{\mu\nu} = \bar g_{\mu\nu} + \delta g_{\mu\nu}$
(denoting for now on the background quantities by an overbar) will read
\begin{equation}
\label{gdg}
\dd s^2 = a^2(\eta)\[-\(1+2A\)\dd\eta^2+2B_i\dd\eta\dd x^i + \(\gamma_{ij}
+h_{ij}\)\dd x^i\dd x^j\],
\end{equation}
whose ``$\varepsilon\to0$'' limit would give (\ref{ConfTime}) back. Note that
the quantity $g^{\mu\nu} = \bar g^{\mu\nu} + \delta g^{\mu\nu}$ should be
the inverse of the above metric, so that demanding $g^{\mu\nu} g_{\nu\alpha}
= \delta^\mu_\alpha$, we obtain $\delta g^{\mu\nu} = - \bar g^{\mu\alpha} 
\bar g^{\nu\beta} \delta g_{\alpha\beta}$.

The stress-energy tensor (\ref{Tmunu}) has background values obtained
with the choice $\bar u^{\mu} = a^{-1}\delta^\mu_\eta$, i.e. 
$\bar u_\mu = -a\delta^\eta_\mu$, and we also demand that the timelike
vector $u^\mu=\bar u^{\mu} + \delta u^\mu$ be normalized at all orders,
leading to $\delta u^\mu = a^{-1} \(-A,v^i\)$, thus defining $v^i$, and
$\delta u_\mu=a\( -A, v_i + B_i\)$; we see that it depends on
the metric perturbation.

Gathering all terms for the stress-energy tensor, we finally obtain
\begin{equation}
\delta T_{\eta\eta} = a^2\rho \(\delta+2A\),\ \ \ \ \delta T_{\eta i} = -a^2
\rho\[\(1+w\)v_i+B_i\] \ \ \ \hbox{and}\ \ \ \  \delta T_{ij} = a^2 p
\( \frac{\delta p}{p}\gamma_{ij} + h_{ij}\),
\label{deltaT}
\end{equation}
where in the last term we have omitted a possible anisotropic stress
contribution. The equation of state itself is perturbed assuming now
that the pressure is a thermodynamical function of both the energy
density and the entropy, if any: $p=p(\rho,S)$. The pressure perturbation
then reads
\begin{equation}
\delta p = \cs^2\delta\rho + \tau\delta S = \cs^2\delta\rho + p\Gamma
=\cs^2\delta\rho + \delta p_\mathrm{nad},
\label{dpdrho}
\end{equation}
where I indicate the most frequently used notations. The last one refers
explicitly to the ``non adiabatic'' component of the pressure, which is
proportional to the entropy variation $\delta S$.

\subsubsection{Scalar, vectors and tensor components}

In the perturbative expansion, we see appearing ordinary functions and
indexed objects. The former transforms as scalars on the spatial hypersurfaces
(recall we have an explicit 3+1 decomposition), while the latter transform
as either vectors or rank-2 tensors of the spatial sections.

Making use of the covariant derivative associated with the metric $\gamma_{ij}$,
which we call $D_i$, one can decompose all relevant quantities in terms of pure
scalar, vector and tensor modes: for instance, the vector $B_i$ appearing
in the metric (\ref{gdg}) can always be written as
\begin{equation}
B_i = D_i B + \hat B_i, \ \ \ \ \hbox{where} \ \ \ \ D^i \hat B_i =0,
\end{equation}
thus exhibiting a scalar function $B$ and two divergenceless vector
degrees of freedom $\hat B_i$, recovering the three initial vector degrees
of freedom. In a more common hydrodynamical framework for instance,
that would be equivalent to splitting the velocity field $v_i$ into a velocity
potential $D_i v$ and a vorticity term $\hat v_i$.

The same technique applies to the tensor quantity $h_{ij}$ which we write
as\footnote{We denote by round parenthesis the symmetrized part of the
relevant tensor, i.e.
$f_{(ij)} \equiv \frac12 \(f_{ij}+f_{ji}\)$.}
\begin{equation}
h_{ij} = 2\[ C \gamma_{ij} + D_{(i} D_{j)} E + D_{(i} \hat E_{j)} + \hat E_{ij}\] \ \ \ \
\hbox{with} \ \ \ \ D^i\hat E_j =0 \ \ \ \hbox{and} \ \ \ D^i\hat E_{ij} = 0 = \hat E^i_{\ j},
\end{equation}
where now the tensor $\hat E_{ij}$ is not only divergenceless but also traceless.
This way, the 10 degrees of freedom of the metric are now split into four scalars
($A$, $B$, $C$ and $E$), 2 vectors ($\hat B_i$ and $\hat E_i$) of 2 degrees of
freedom each, and one tensor $\hat E_{ij}$, also having 2 independent degrees
of freedom. The main interest of this Scalar-Vector-Tensor (SVT)
decomposition is that, at linear order,
they all decouple, and one can thus treat the scalar, vector and tensor modes
independently. 

\subsection{The gauge issue}

GR is diffeomorphism invariant, i.e. it is constructed in such a way that
general coordinate transformations leave the equations unchanged. This
implies that out of the 10 degrees of freedom discussed above, 4 are
essentially irrelevant as they can be absorbed into a coordinate
transformation. When applied
to the special background $+$ perturbations case, this invariance is no longer
an actual coordinate transformations since the background is kept fixed;
it is then called a {\sl gauge transformation}.
Let us see in more details how
it works.

\subsubsection{Metric fluctuations}

Suppose I change the coordinates $x^\mu$ to a set of new coordinates
$\tilde x^\mu$ related with the previous ones by an infinitesimal translation, i.e.
$x^\mu \mapsto \tilde x^\mu = x^\mu + \xi^\mu$, where $\xi^\mu$ are small
quantities. General covariance then implies that the equations of motion have
the same form when expressed in the ``new'' coordinates $\tilde x^\mu$ or
the ``old'' ones $x^\mu$. In particular, the line element, namely
$\dd s^2$, should have the same structure under a gauge transformation. Therefore,
we set
\begin{equation}
\label{gdgtilde}
\dd \tilde s^2 = a^2(\eta)\[-\(1+2\tilde A\)\dd\tilde\eta^2+2\tilde B_i\dd\tilde
\eta\dd \tilde x^i + \(\gamma_{ij}
+\tilde h_{ij}\)\dd \tilde x^i\dd \tilde x^j\],
\end{equation}
and by gauge invariance, we require that $\dd \tilde s^2 = \dd s^2$,
after having SVT-decomposed the transformation through
$\tilde\eta = \eta +T$ and $\tilde x^i = x^i +D^i L +\hat L^i$. We find
the following transformation laws:
\begin{equation}
\tilde A = A -\( T'+\Hu T\),\ \ \ \ \tilde B = B-\(L'-T\),\ \ \ \ \tilde C = C-\Hu T
\ \ \ \ \hbox{and} \ \ \ \ \ \tilde E = E-L
\label{scalarTr}
\end{equation}
for the scalar quantities, 
\begin{equation}
\hat{\tilde B}^i = \hat B^i- \bar L^{i\prime} \ \ \ \hbox{and} \ \ \ \  \hat{\tilde E}^i = \hat E^i
- \bar L^i
\label{vectorTr}
\end{equation}
for the vectors and finally $\hat{\tilde E}_{ij} = \hat E_{ij}$. The last
identity could have been obtained without any calculation from
the vectorial nature of the transformation: tensor modes, also called
gravitational waves, are naturally gauge invariant.

Equation (\ref{vectorTr}) can easily be reshuffled into $\hat{\tilde E}^{i\prime} - \hat{\tilde B}^i =
\hat E^{i\prime} - \hat B^i$, so that the quantity $\bar \Phi^i \equiv
\hat E^{i\prime} - \hat B^i$ is gauge invariant. On the scalar side, similarly, one
finds that $\(\tilde B-\tilde E'\) = \(B-E'\)+T$, so that $\[\tilde A +\(\tilde B-\tilde E'\)\]
=\[A +\(B-E'\)\]-\Hu T$, and finally that the quantity 
\begin{equation}
\Phi \equiv A+\(B-E'\)'+\Hu \(B-E'\)
\label{Phi}
\end{equation}
is also gauge invariant. I leave it as an exercise to show that
\begin{equation}
\Psi \equiv -C-\Hu\(B-E'\)
\label{Psi}
\end{equation}
closes the set of gauge-invariant variables consisting of two scalars $\Phi$ and $\Psi$,
called the Bardeen potentials,
two vectors $\hat\Phi^i$ and two tensors $\hat E_{ij}$ for a total of 6 gauge-invariant
quantities, as expected from the original 10 quantities and 4 possible gauge
choices.

\subsubsection{Choosing a gauge}

One can play the same game with the stress-energy tensor and
obtain transformation rules by expressing it in one frame or the
other through the usual transformation rule of a rank-2 tensor. One
finds
\begin{equation}
\widetilde{\delta\rho} = \delta\rho +\rho'T, \ \ \ \ \tilde v = v-L', \ \ \ \ \hat{\tilde v}^i = \hat v^i
-\hat L^{i\prime} \ \ \ \hbox{and} \ \ \ \widetilde{\delta p} = \delta p + p'T,
\end{equation}
leading here also to a set of gauge-invariant variables
\begin{equation}
\delta\rho^\mathrm{_N} \equiv \delta \rho+ \rho' \(B-E'\), \ \ \ \ 
\delta p^\mathrm{_N} \equiv \delta p+ p'\(B-E'\), 
\ \ \ V\equiv v+E'\ \ \ \ \hbox{and} \ \ \ \ \bar V^i \equiv \bar v^i + \bar B^i,
\label{Newton}
\end{equation}
given here an only one example of such a combination.

{}From that point on, one can write down Einstein equations and solve them: just like
in electromagnetism, one merely needs to fix a gauge. There are many gauges that
have been used in the literature, and I list a few of them here. The first I want to list
shows that the gauge-fixing choice is, just like in electromagnetism again, not necessarily
enough: it is the so-called synchronous gauge, in which only spatial sections are
perturbed. In other words, it is defined by assuming that the proper time of a comoving
observer is cosmic time, and this translates into setting $A=0$ and $B_i=0$. Because
of its definition, it is a quite intuitive gauge, but it is not completely fixed: setting
$\tilde \eta = f(\eta)$ or $\tilde x^i = f^i\( x^j\)$ for arbitrary fonctions $f$ and $f^i$,
one remains in this gauge ($\tilde A=0$ and $\tilde B_i=0$ are still valid). This leads
to possibly spurious solutions, and hence to mistakes!

Another frequently used gauge in the case of a single fluid is one which follows the
fluid's motion, so that one demands $\delta T^0_{\ i}=0$. This is an interesting choice
which becomes unfortunately ambiguous as soon as more than one fluid is involved.
In this gauge, the variables
\begin{equation}
\delta\rho^\mathrm{_C} \equiv \delta\rho + \rho' \(v+B\)\ \ \ \hbox{and} \ \ \ 
\delta p^\mathrm{_C} \equiv \delta p+ p'\(v+B\),
\label{comob}
\end{equation}
are the natural fluid variables to use.

Another physically interesting choice is that which consists in demanding the curvature
perturbation of spatial section to vanish, which amounts to setting $C=E=0$ and
$\hat E_i=0$, so the quantities
\begin{equation}
\delta\rho^\mathrm{_F} \equiv \delta\rho - \rho' \frac{C}{\Hu} \ \ \ \hbox{and} \ \ \ 
\delta p^\mathrm{_F} \equiv \delta p- p'\frac{C}{\Hu} ,
\end{equation}
reduce to their original values: these gauge-invariant variables are thus
the density and pressure perturbations in the flat-slicing gauge.

Finally, it seems also appropriate to use directly a set of physically relevant variables
like those defined above, namely the gauge-invariant ones. The simplest way to do
that is to impose the so-called longitudinal, or Newtonian, gauge, i.e. that in which
the scalar part of $g_{\mu\nu}$ is diagonal so that we set $E=B=0$. In this gauge,
the potential (\ref{Psi}) is the Newtonian potential. There are two possibilities to
get to this gauge: one can either set $E=B=0$ from the outset (easy way) or work
out all the equations and express all of them only in terms of the gauge-invariant
variables (\ref{Phi}) and (\ref{Psi}). They both give the exact same results, of
course. One sees that in this gauge, the density and pressure perturbations reduce
naturally to those defined in (\ref{Newton}).

\subsubsection{Perturbed Einstein equations}

We now are in a position to write down explicitly the Einstein equations
to first order of perturbations in a meaningful way. The equations in the
Newtonian gauge only involve gauge-invariant quantities, and I shall
therefore restrict attention to those in what follows. Since the following
section is dedicated to tensor modes, I will simply forget about them
until then (remember they decouple at linear order anyway).

The next-to-simple case is that of vector modes. In most cosmologically
relevant situations, there is no anisotropic stress ($\hat\pi_i=0$), so that
the equations
of motion of the vector modes are not sourced by anything. They take the
form
\begin{equation}
\(\Delta + 2 \Ka\) \hat\Phi_i = -\frac{16\pi\GN}{3} \rho a^2 \(1+w\) \hat V_i,
\label{Vi}
\end{equation}
and, more importantly
\begin{equation}
\hat \Phi_i' + 2\Hu \hat \Phi_i = \frac{8\pi\GN}{3} p a^2 \hat \pi_i \to 0,
\label{Phii}
\end{equation}
leading to the exact solution $\hat \Phi_i \propto a^{-2}$, and consequently,
thanks to (\ref{Vi}), that $\hat V_i \propto a^{3w-1}$. It is a well-known
(observational) fact
that vector modes were negligible at the time of nucleosynthesis, so we may
confidently set $||\hat \Phi_i||\ll 1$ at $z_\mathrm{nucl} \sim 3\times 10^8$. This
implies $||\hat \Phi_i||\ll 10^{-17}$ now: apart in very special situations such as
a contracting universe in a bouncing scenario, one can set the vector
perturbation to zero. I shall not consider them anymore in what follows.

We are thus left with scalar modes. Since those have been driving the
gravitational collapse leading to large-scale-structure formation, they are
definitely the most relevant modes to study and, indeed, they play the
first role in most of the literature on the subject. Their time development
is obtained through two independent sets of equations, the first relating
density to pressure perturbations, i.e. Eq.~(\ref{dpdrho}), the rest being
given by Einstein equations, the spatial part of which, proportional to
$\delta T^i_{\, j}\propto \delta^i_{\, j}$ for a perfect fluid, yielding $\gamma^{ij}
D_i D_j \( \Phi-\Psi\) = 0$: under the reasonable assumption that the
scalar perturbations do not diverge at spatial infinity, this relation implies
that the only possibility is to have $\Psi=\Phi$, a condition which I will take
as valid for now on.

For the scalar modes in the longitudinal gauge, Einstein equations then read
\begin{align}
& \Delta \Phi  - 3\Hu \Phi' -3\( \Hu^2-\Ka\) \Phi = 4\pi\GN a^2 \delta\rho^\mathrm{_N},
\label{DPhi} \\
& D_i \(\Phi'+\Hu \Phi\) = -4\pi\GN a^2 \( \rho+p\) \nabla_i V,
\label{DiPhi}\\
& \Phi'' +3\Hu\Phi' +\(2\Hu'+\Hu^2-\Ka\) \Phi = 4\pi \GN a^2 \delta\rho^\mathrm{_N}.
\label{Phipp}
\end{align}
Equation (\ref{DPhi}) can be reformulated as $\( \Delta +3\Ka\) \Phi = 4\pi\GN a^2 \delta\rho^\mathrm{_C}$,
using (\ref{comob}) and (\ref{DiPhi}). This Poisson equation (up to the spatial curvature term)
shows that the Bardeen potential is essentially the ordinary Newton potential if the density
perturbation is expressed in the comoving gauge. As we shall see later, the matter perturbations
in the different gauges on scales smaller than the Hubble radius are basically the same, so the
sub-Hubble Bardeen potential indeed reduces to the Newtonian one (hence the notation $\Phi$).

Now, using Eq.~(\ref{dpdrho}) to express the pressure perturbation in terms of the density,
and then replacing (\ref{Phipp}) into (\ref{DPhi}), one obtains
\begin{equation}
\Phi''+3\Hu\(1+\cs^2\)\Phi'-\cs^2\Delta\Phi +\[2\Hu'+\(1+3\cs^2\)\(\Hu^2-\Ka\)\]\Phi
= 4\pi\GN a^2 \tau \delta S,
\label{PhiEvol}
\end{equation}
which can be understood as the general relativistic version of Eq.~(\ref{pertGen}).

Finally, this evolution equation can be made to a much simpler, intuitive
and tractable form: by setting
\begin{equation}
\label{udef}
u=\frac43 \frac{a^2\theta}{\Hu} \Phi \ \ \ \hbox{with} \ \ \ \ \theta\equiv \sqrt{\frac{3}{2a^2\Gamma}}
\ \ \ \hbox{and}\ \ \ \Gamma\equiv 1-\frac{\Hu'}{\Hu^2} + \frac{\Ka}{\Hu^2},
\end{equation}
one can check after a few tedious but straightforward calculation that Eq.~(\ref{PhiEvol})
takes the wavelike form
\begin{equation}
\label{ueq}
\boxed{u''+\( \cs^2 k^2 -\frac{\theta''}{\theta} \) u = \frac{8\pi\GN}{3} \frac{a^4 \theta}{\Hu} \tau \delta S,}
\end{equation}
where I have replaced the Laplacian $\Delta\to -k^2$ in Fourier space. When there is
no entropy perturbation (adiabatic perturbations), this equation is simply that of a parametric
oscillator; the entropy contribution can then be seen as a forcing term. As it turns out to be the same as the gravitational wave case,
I now move to those.

\section{The example of tensor modes}\label{sec:tensor}

Tensor modes, being gauge-invariant from the outset, are free of all
gauge-fixing subtleties, and can be computed straightforwardly. Einstein
equations for those read
\begin{equation}
\label{tensmod}
\hat E_{kl}'' + 2\Hu \hat E_{kl}' + \(2\Ka -\Delta\)\hat E_{kl} = 
8\pi\GN a^2 p \hat \pi_{kl},
\end{equation}
where the anisotropic stress $\hat\pi_{kl}$ is usually set to zero, in
agreement with the observations. Moreover, as we have seen, the
spatial section curvature is also measured to be quite small, so we
can safely set it to zero as well. Since inflation also set both these
quantities to vanishingly (exponentially) small values, we have both
observational and theoretical good reasons to set $\hat\pi_{kl}\to 0$
and $\Ka\to 0$.

\subsection{Flat space polarization}

In order to understand what a tensor mode is, it is simpler to first
consider the non expanding case in which we set $\Hu\to 0$, so the
Einstein equation for $\hat E_{ij}$ reduces to the wave equation
\begin{equation}
\Box \hat E_{ij} = 0
\label{wave0}
\end{equation}
whose solutions I now discuss.

\subsubsection{Polarization.}

Let us consider for simplicity a mode propagating along the $z$
direction, and pick the simplest possible solution of (\ref{wave0}), i.e.
$\hat E_{ij}\propto \cos\[k(z-t)\]$. Now what is missing in this
solution is the set of indices, which account for the polarizations. With
$k_i=(0,0,k)$, the transverse and traceless conditions for $\hat E_{ij}$
read $\hat E_{xz} = \hat E_{yz} = \hat E_{zz} = 0$, $\hat E_{xx}
=\hat E_{yy}$ and $\hat E_{xy} = \hat E_{yx}$. We are thus left
with two independent solution, $\hat E_{xx}$ and $\hat E_{xy}$ say.
These are the functions behaving as sines and cosines.

\begin{figure}[h]
\centering
\includegraphics[width=15.0cm,clip]{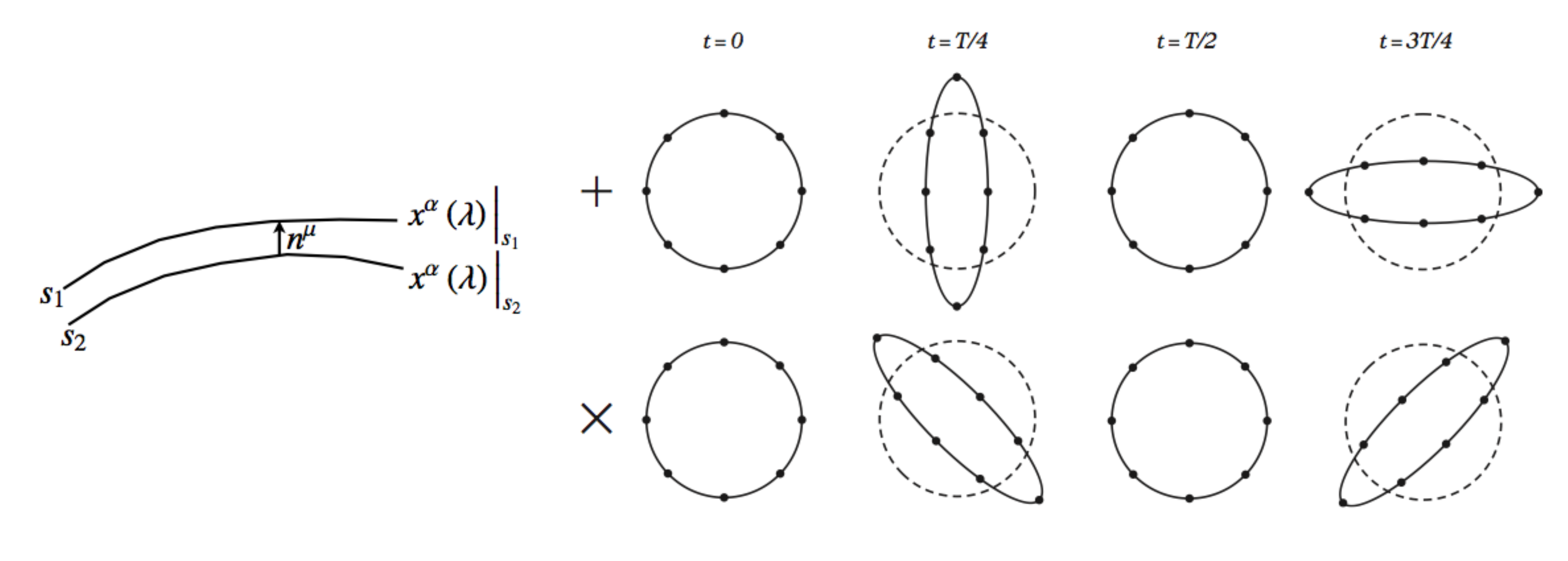}
\caption{Left: Geodesic deviation $n^\mu$ of two geodesics
defined by $x^\mu(\lambda)\big|_{s_1}$ and $x^\mu(\lambda)\big|_{s_2}$
and representing the trajectories of two test particles. Right: When a
gravitational wave mode $\varepsilon^+$ or $\varepsilon^\times$
passes through a ring of such test particles, they evolve as
shown, producing the `$+$' or `$\times$' shapes with time, for
different values of time in units of the period $T$.}
\label{Xplus}
\end{figure}

The full solution can be expressed in terms of these functions together
with a set of polarization tensors $\varepsilon^+_{ij}$ and 
$\varepsilon^\times_{ij}$, namely
\begin{equation}
\label{solModes}
\hat E_{ij} = \( \begin{array}{ccc}
\hat E_{xx} & \hat E_{xy} & 0
\\ \hat E_{xy} & -\hat E_{xx} & 0\\
0&0&0
\end{array}
\) = \underbrace{
 \( \begin{array}{ccc}
1 & 0 & 0
\\ 0 & -1 & 0\\
0&0&0
\end{array}
\)}_{\varepsilon^+_{ij}}
\hat E_{xx} \(\bm{x},t\) 
+ \underbrace{\( \begin{array}{ccc}
0 & 1 & 0
\\ 1 & 0 & 0\\
0&0&0
\end{array}
\)}_{\varepsilon^\times_{ij}} 
\hat E_{xy}\(\bm{x},t\) 
\end{equation}
whose names stem from their effect on a test particle.

\subsubsection{Observing a gravitational wave}

Let us consider two such neighboring test particles following their
own paths $x^\mu(\lambda)\big|_{s_1}$ and $x^\mu(\lambda)\big|_{s_2}$
(see Fig. \ref{Xplus}) and assume they are originally at rest. The 
connections $\Gamma^i_{\ 00}$ being vanishing at first order in perturbations
(only $\hat E_{ij}$ is present), as the wave passes, a particle originally at
rest remains apparently so: the particle is moving with the reference
frame. However, the perturbed curvature is non vanishing, so the relative
geodesic motion is affected by the wave. The geodesic deviation $n^\mu=\partial
x^\mu/\partial s$ between these geodesics feels an acceleration given by
\begin{equation}
a^\mu = \frac{\dd^2 n^\mu}{\dd \lambda^2} = u^\alpha\nabla_\alpha\( u^\beta
\nabla_\beta n^\mu \) = R^\mu_{\ \nu\alpha\beta}u^\nu u^\alpha n^\beta,
\end{equation}
so that the distance between the ring-forming particles changes with
time as
$$
\frac{\dd^2 n^i}{\dd t^2} = \frac12 \partial_t^2 \hat E_{j}^{\, i} n^j,
$$
leading to the time evolutions shown in Fig. \ref{Xplus}.

The very simple cosine and sine solutions are obtained in the
flat Minkowski case, and can readily be generalized to the
expanding case with a scale factor increasing as a power law:
they are then replaced by Bessel functions, see below.

\subsection{Cosmological gravitational waves}

\begin{figure}[h]
\centering
\includegraphics[width=12.0cm,clip]{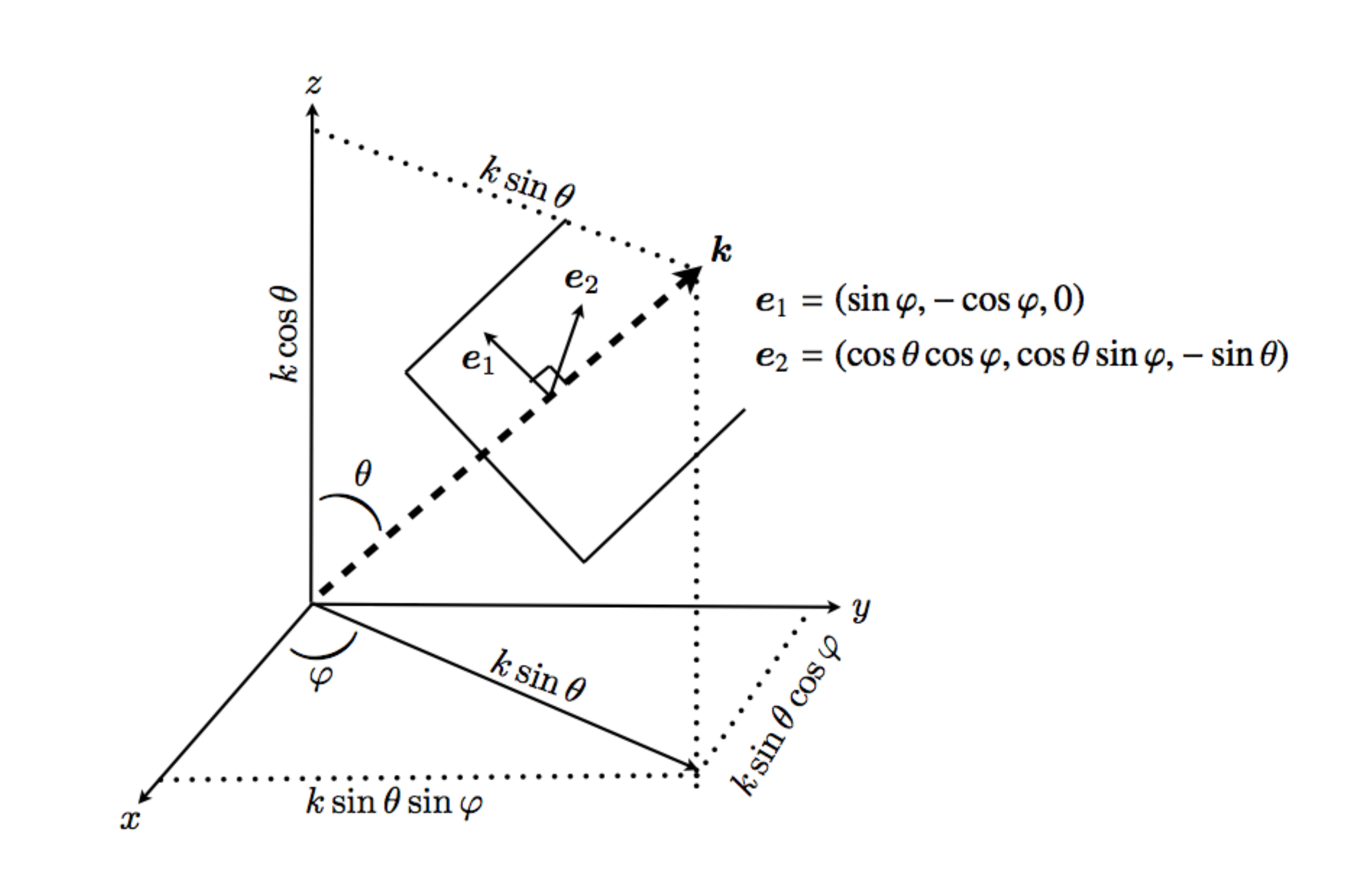}
\caption{Notations for the gravitational wave propagation: the wave propagates
along the direction of $\bm{k}$, and the dyad $\bm{e}_a$ ($a=1,2$), in the
plane orthogonal to $\bm{k}$, is used to define the polarization tensor.}
\label{Polarisations}
\end{figure}

Even under the simplifying assumptions made below
Eq.~(\ref{tensmod}), one still needs take into account
the  tensorial nature of the modes. Because this equation
is linear, it can easily be decomposed into Fourier modes in the form
\begin{equation}
\hat E_{ij} = \frac{1}{a}\frac{1}{\(2\pi\)^{3/2}} \sum_{\lambda=1}^2 \int
\dd^3\bm{k}\, P^\lambda_{ij}\(\bm{k}\) \mu_\lambda \(\eta,\bm{k}\)
\ex^{i\bm{k}\cdot \bm{x}},
\label{FourierTens}
\end{equation}
where $\lambda$ is a polarization index (running from 1 to 2 because
of the two degrees of freedom in the tensor modes), and the
polarization tensor $P^\lambda_{ij}$ can be given explicitly as follows.
Note the factor $1/a$ which has been put here for further
convenience.

\subsubsection{Polarization modes}

Figure \ref{Polarisations} displays the configuration for the wave
vector $\bm{k}$ and the orthogonal plane in which one defines
a dyad $\bm{e}_a$ ($a=1,2$) satisfying $\bm{e}_a\cdot \bm{e}_b
=\delta_{ab}$ and $\bm{k}\cdot\bm{e}_a=0$. Recall that the
tensor mode $\hat E_{ij}$ is transverse and traceless, which
translates into the polarization tensor as the requirement
\begin{equation}
k^i P^\lambda_{ij} = 0,\ \ \ \ \hbox{and} \ \ \ \ P^\lambda_{ij} \delta^{ij} = 0,
\label{tensconst}
\end{equation}
(we have set $\Ka\to 0$ and thus can identify $\gamma^{ij}
\to \delta^{ij}$). One can check that the choice
\begin{equation}
P^{(1)}_{ij} = \(\bm{e}_1\)_i\(\bm{e}_1\)_j - \(\bm{e}_2\)_i\(\bm{e}_2\)_j
\ \ \ \ \hbox{and} \ \ \ \ P^{(2)}_{ij} = \(\bm{e}_1\)_i\(\bm{e}_2\)_j+
\(\bm{e}_2\)_i\(\bm{e}_1\)_j
\label{Pij}
\end{equation}
satisfies all the constraints (\ref{tensconst}).

Plugging the form (\ref{Pij}) into Eq. (\ref{tensmod}) then shows that
both quantities $\mu_\lambda$ satisfy the same differential equation. We
then simply set $\mu_{\lambda=1,2} \equiv \mu_{_\mathrm{T}}$ (the index
`T' standing for tensor, we will later have a similar variable with an
index `S' for the scalar case) and obtain
\begin{equation}
\label{muT}
\boxed{\mu_{_\mathrm{T}}''+\( k^2 -\frac{a''}{a}\) \mu_{_\mathrm{T}} =0,}
\end{equation}
which is the prototypical wavelike equation obtained in cosmological
perturbation theory. It is interesting to realize that it can be obtained
by varying the Einstein-Hilbert action expanded to second order
in perturbation, namely (for the general case including curvature)
\begin{equation}
\delta^{(2)}S_{_\mathrm{T}} = \frac12 \sum_{\lambda=1}^2\int\dd^3\bm{x} \dd \eta
\sqrt{\gamma} \[ \(\mu_\lambda'\)^2 - \gamma^{ij}\partial_i\mu_\lambda \partial_j\mu_\lambda +
\(\frac{a''}{a}-2\Ka\)\mu_\lambda^2 \],
\label{GravMod2}
\end{equation}
in which one recognizes the action of a time varying mass scalar field. This
observation lies at the heart of the idea of setting quantum initial conditions,
as I will explain later.

\subsubsection{Time development of a mode}

\begin{figure}[h]
\centering
\includegraphics[width=15.0cm,clip]{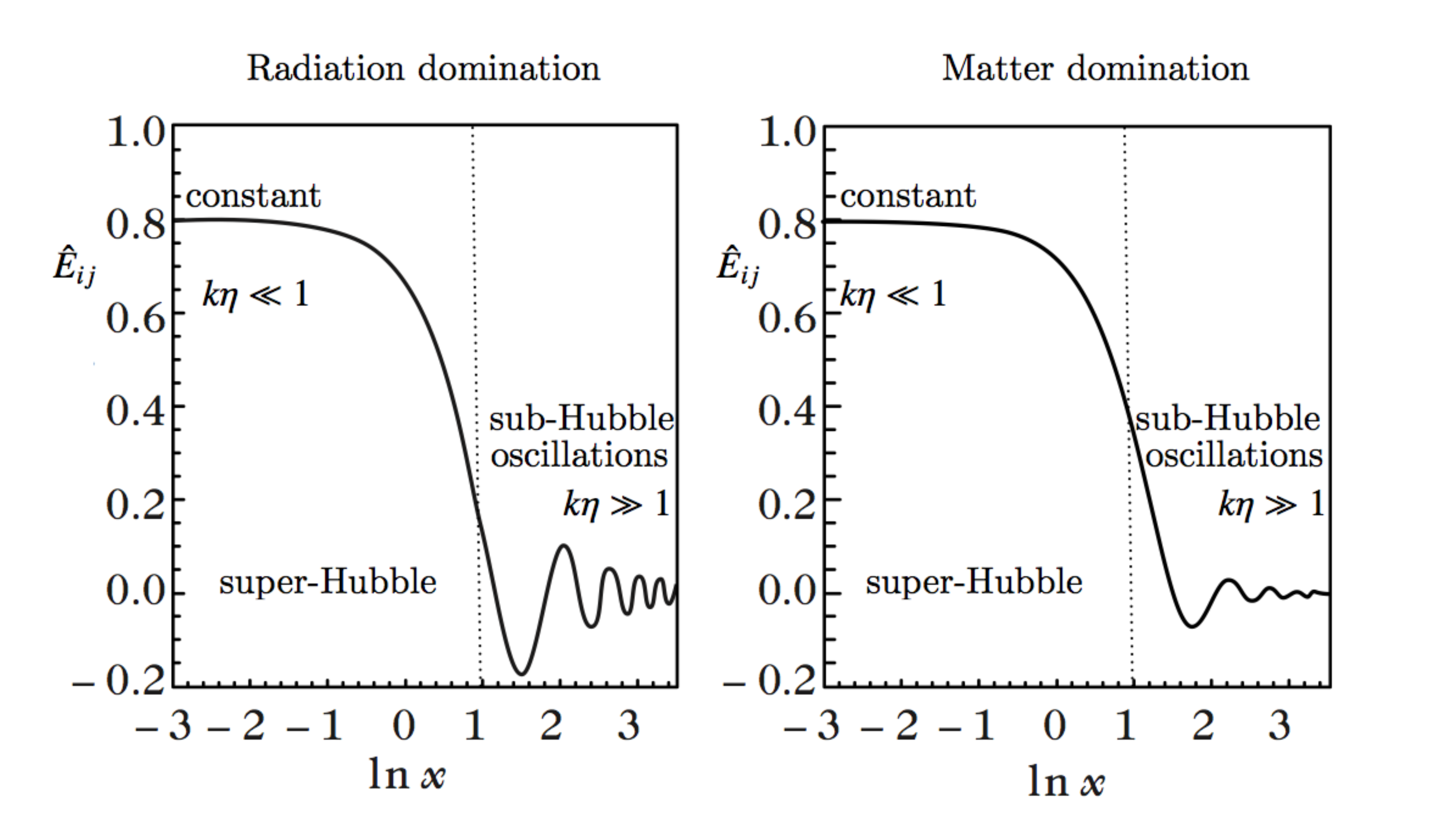}
\caption{Solutions of the gravitational wave modes equation
(\ref{Bessel}) for a given polarization mode $\hat E_{ij}$ as
a function of $x\equiv k\eta$ for radiation ($\nu=1$, left
panel) and matter ($\nu=2$, right panel) domination. The
dotted line represent the Hubble crossing time $k\eta=1$:
for $k\eta\ll 1$, i.e. deep in the super-Hubble regime, the
solution behaves as a constant, while for the sub-Hubble
case $k\eta\gg 1$, the solution exhibits the oscillatory behavior
already encountered in the Minkowski case.}
\label{SolBessel}
\end{figure}

Let us go back to the original equation (\ref{tensmod}) in the case of
an expanding Universe dominated by a perfect fluid with constant
equation of state. In this case, we have seen that the scale factor
behaves as $a\propto \eta^\nu$ for some value of $\nu$, and the
Hubble function then takes the simple form $\Hu = \nu/\eta$. 
Eq.~(\ref{tensmod}) thus becomes
\begin{equation}
\frac{\dd\hat E_{ij}}{\dd x^2} + \frac{2\nu}{x} \frac{\dd \hat E_{ij}}{\dd  x}
+\hat E_{ij}=0,
\label{Bessel}
\end{equation}
where $x\equiv k\eta$. As announced earlier, this is a Bessel
equation whose solutions have been studied in details since
the beginning of the 19$^\mathrm{th}$ century. They are shown
on Fig. \ref{SolBessel} and read
\begin{equation}
\hat E_{ij} = x^{\frac12-\nu}\[ A_{ij} J_{\nu-\frac12} \(x\) +
B_{ij} N_{\nu-\frac12} \(x\)\],
\label{BesselFuncts}
\end{equation}
where the tensors $A_{ij}$ and $B_{ij}$ are, of course, transverse
and traceless.

The solutions show two extreme regimes, called sub-
and super-Hubble. They refer to the characteristic
ratio of the wavelength ($k$) to the Hubble scale
($\propto\eta^{-1}$ for a power-law scale factor), i.e. the
variable $x=k\eta$. Long wavelengths ($x\ll 1$) 
are strongly damped
by the expansion, and so any initial motion is rapidly erased
by the expansion-induced friction and the mode behaves as
a constant. For short wavelengths ($x\gg 1$)
on the other hand, the
expansion is negligible and the mode behaves essentially as in
a Minkowski universe: we recover the oscillations obtained
earlier.

\subsubsection{Hubble vs Horizon}

At this stage, I think it is important to make a short comment
on a commonly used and very misleading phrase, namely the
use of ``sub-horizon'' and ``super-horizon'' modes instead
of sub-Hubble and super-Hubble. As shown by Eq. (\ref{hor}),
the horizon is a global quantity which depends on the
entire history of the Universe. Early models were based on
the singular big-bang followed by radiation and matter dominations,
and it is easy to convince oneself that in this case, the integrated
quantity $d_{_\mathrm{H}}$ is, up to an irrelevant numerical
factor, given by the Hubble scale. In such a context, comparing
wavelengths with the Hubble radius would indeed be equivalent
to comparing them to the size of the horizon... but it is then not
very clear what the meaning of these would have been! Indeed,
in GR, modes larger than the horizon are actually not well-defined
in the sense that setting initial conditions for them would be
a direct violation of causality.

In fact, one often reads that the modes are ``frozen'' because
of some ``causality'' reason, with the meaning that a mode larger
than the horizon could not evolve at all because of causality,
as both ends of the mode would need to propagate faster than
light to communicate, which is forbidden. I do not know what is
the meaning of such arguments, and strongly suspect they have
none whatsoever. A given mode
consists of a linear combination of the two independent functions
solving a second order linear equation\footnote{This discussion is also valid
in the case of scalar perturbations, so I do not restrict attention here
to the tensor case.}, with coefficients provided by the initial
conditions. Then, for a super-Hubble wavelength, what happens
is that the expansion very rapidly suppresses one of the solution relative
to the other, and one is left with the constant mode as discussed above.
But this is in no way related with causality, on the contrary, it is a
purely dynamical statement.

We have discussed this point in more details in Ref. \cite{JMPPPRL}
for the specific case of scalar perturbations in bouncing models.

\subsubsection{Radiation-to-matter transition}

With the mode evolution known in any given epoch and a primordial
spectrum, one should in principle be able to predict the observed spectrum.
As it turns out, the theories that agree with the data predict an almost scale-invariant
initial spectrum, i.e. one in which no particular scale is singled out. On the
other hand, we know that such a scale should be present somehow, because
the Universe, which was at very early times dominated by radiation, transitioned
to the matter era\footnote{There was another transition more recently when
the Universe became dominated by the cosmological constant or whatever
it is which mimics it nowadays; I will not discuss this any further, but in
principle, it could well lead to another scale in the data indeed.}.

The transition can be treated simply by introducing a new variable
$y\equiv a/a_\mathrm{eq}$, where $a_\mathrm{eq}$ is the value of
the scale factor at equality between radiation and matter, shown in
Fig. \ref{rhos} and defined through $\rho_\mathrm{m}
\(a_\mathrm{eq}\) = \rho_\mathrm{r} \(a_\mathrm{eq}\)$. Given
that $\rho_\mathrm{m}= \rho_\mathrm{m}^0 a^{-3}$ and $\rho_\mathrm{r}
=\rho_\mathrm{r}^0 a^{-4}$, we find that $a_\mathrm{eq} = \rho_\mathrm{m}^0
/\rho_\mathrm{r}^0$, and finally that $y=\rho_\mathrm{m}/\rho_\mathrm{r}$.
I leave as an exercise to the reader to show that the total equation
of state $w$, defined as the ratio of the total pressure by the total energy
density, is $w=\frac13 \(1+y\)^{-1}$ and the sound velocity is
$\cs^2 = \frac13  \(1+\frac34 y\)^{-1}$.

\begin{figure}[h]
\centering
\includegraphics[width=13.0cm,clip]{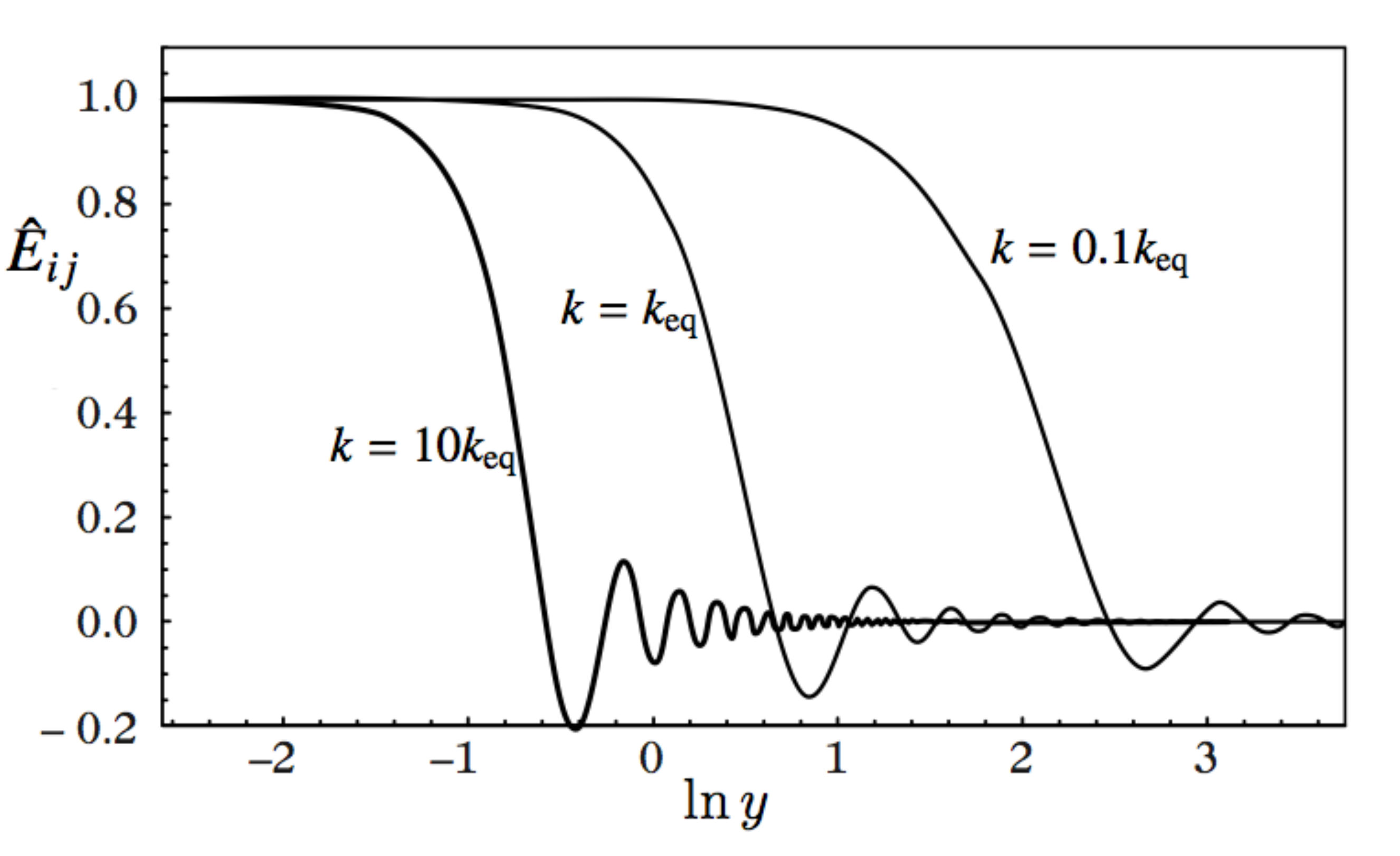}
\caption{Solutions of the gravitational wave modes equation
(\ref{ModeTensTrans}) during the radiation-to-matter
transition for 3 different values of the wavenumber $k$ as
function of the rescaled scale factor variable $y=a/a_\mathrm{eq}$.
Long wavelength modes (small $k$) see the transition later than
short wavelength modes (large $k$): the latter are more damped than
the former, and the characteristic scale of the transition is thus
imprinted into the subsequent spectrum.}
\label{SolE}
\end{figure}

The Friedmann equation (\ref{calH2}) takes the form
\begin{equation}
\Hu^2 = \frac{8\pi\GN}{3} \rho a^2 = \frac{8\pi\GN}{3} \rho_\mathrm{r} 
\(1+y\) y^2 a_\mathrm{eq}^2 \ \ \ \Longrightarrow \ \ \ \Hu_\mathrm{eq}^2
= \frac{16\pi\GN}{3}\frac{\rho_\mathrm{r}^0}{a_\mathrm{eq}^2} \ \ \ \Longrightarrow
\ \ \ \Hu^2 = \frac{1+y}{2 y^2} \Hu_\mathrm{eq}^2,
\label{FriedTrans}
\end{equation}
thus allowing to switch to the variable $y$ whenever one encounters $\Hu$.
Noting that the derivatives with respect to $\eta$ and $y$ satisfy
$$\frac{\dd}{\dd\eta} = \frac{\dd y}{\dd\eta} \frac{\dd}{\dd y}=\frac{a'}{a_\mathrm{eq}}
\frac{\dd}{\dd y}=
\Hu y \frac{\dd}{\dd y},$$
and defining the wavenumber characteristic of equality as $k_\mathrm{eq} 
= \Hu_\mathrm{eq}= a_\mathrm{eq}H_\mathrm{eq}$, we find that Eq. (\ref{tensmod})
takes the following form
\begin{equation}
\frac{\dd^2\hat E_{ij}}{\dd y^2} + \frac{4+5y}{2y(1+y)} \frac{\dd\hat E_{ij}}{\dd y} 
+ \( \frac{k}{k_\mathrm{eq}}\)^2 \frac{2}{1+y} \hat E_{ij} = 0,
\label{ModeTensTrans}
\end{equation}
whose analytic solution is not known... but we can solve it numerically for
different values of $k$. This is done in Fig. \ref{SolE}.

Let me conclude this section by returning to the form (\ref{muT}) of the
mode equation. Its formal solution is known and given by the Born
expansion, namely
\begin{equation}
\mu_{_\mathrm{T}}\(k,\eta\) = a(\eta) \[ A_1(k) +A_2(k) \int^\eta \frac{\dd\tau}{a^2(\tau)}
+ k^2 \int^\eta\frac{\dd\sigma}{a^2(\sigma)}\int^\sigma\dd\tau a^2(\tau) \mu_{_\mathrm{T}}
(k,\tau)\],
\label{Born}
\end{equation}
i.e., we have replaced a differential equation by an integral one! However, we
have gained something in the process because the solution is known exactly
in the long and short wavelength limits. They are, for short scales
\begin{equation}
k^2\gg \frac{a''}{a} \ \ \ \Longrightarrow \ \ \ \mu_{_\mathrm{T}} = A(k) \ex^{ik\eta}
+ B(k) \ex^{-ik\eta},
\label{short}
\end{equation}
and for the large scales
\begin{equation}
k^2\ll \frac{a''}{a} \ \ \ \Longrightarrow \ \ \ \mu_{_\mathrm{T}} = C(k) a
+ D(k) a\int^\eta \frac{\dd \tau}{a^2(\tau)},
\label{long}
\end{equation}
so that it suffices to plug one of these solution into (\ref{Born}) to
obtain an expansion in either large or short wavenumbers.

One final point, regarding the action (\ref{GravMod2}) and the 
solution (\ref{short}) valid for small wavelengths, i.e. when the
expansion can be discarded: in this case, we can consider
$\mu_{_\mathrm{T}}$ as an actual scalar field in a Minkowski
universe, a system which we know how to quantize. Once the
field itself is quantized, one can impose a given physical
quantum state, for instance the vacuum. But this actually
fixes unambiguously the coefficients $A(k)$ and $B(k)$... and
then, the theory becomes predictive! I shall return to this in
the last part of these lectures.

After having discussed the tensor modes in quite some depth, let
me now sketch the scalar case which, although currently the only
one observed, is also sufficiently more involved to require a special
treatment of its own. This is done in Refs. \cite{PPJPU} and
\cite{Mukhanov:2005sc}.

\section{Density fluctuations and the power spectrum}

Since the tensor modes have not been observed yet, let us move on
to the scalar case. For this, I will set a much oversimplified (and
already disproved by the data!) model having $\Ka=0$, i.e. $\sum_i
\Omega_a\equiv\Omega_\mathrm{tot}=1$, and assume all
there is to consist of matter now (i.e. $\Omega^0_\mathrm{m}=1$)
with a currently negligible amount of radiation $\Omega^0_\mathrm{r}\ll1$.

\subsection{Basic equations}

Before we move on to our specific example involving basically
only matter ($w_\mathrm{m}$) and radiation ($w\dr$), let us write
down the more general set of equations
for a fluid having $N$ constituents, e.g. radiation, dust, neutrinos,
dark energy, dark matter, and whatever else a theoretician's brain
can come up with.

The total energy density $\rho = \sum_a\rho_a$ and pressure
$p=\sum_a p_a$ allow to define a global velocity through
$\(\rho+p\) v^i = \sum_a\(\rho_a + p_a\) v^i_a$. The total
equation of state $w\equiv p/\rho$ and sound velocity $\cs^2
\equiv p'/\rho'$ can be obtained as
\begin{equation}
\Omega w = \sum_a \Omega_a w_a \ \ \ \ \hbox{and} \ \ \ \ 
\Omega \cs^2 = \sum_a \frac{1+w_a}{1+w} \Omega_a c_a^2,
\label{wcs}
\end{equation}
where each fluid sound speed is $c_a^2\equiv p_a'/\rho_a'$.

Similar calculations can be made at the perturbation level,
yielding $\Omega\delta = \sum_a\Omega_a\delta_a$ for
the density fluctuations and $\(1+w\) \Omega v=\sum_a
\(1+w_a\) \Omega_a v_a$ for the velocities. The total
entropy perturbation can be derived in much the same
way as for the single fluid case, namely recalling that we
set $\tau\delta S= P\Gamma$, we get $w\Gamma =
\(\delta p-\cs^2 \delta\rho\)/\rho$, and finally
\begin{equation}
\Omega w \Gamma = \sum_a w_a\Gamma_a + \sum_a \Omega_a
\delta_a \(c_a^2-\cs^2\),
\label{Gamma}
\end{equation}
showing that even if each individual fluid has vanishing
self entropy perturbation (i.e. even if all $\Gamma_a\to0$), the
total fluid mixing entropy can be non vanishing.

In principle, if the fluids are coupled, one should not necessarily assume
them to be each independently conserved, but rather to satisfy
(the condition in parenthesis being redundant)
\begin{equation}
\nabla_\mu T^{\mu\nu}_a = Q^\nu_a \ \ \ \ \hbox{with} 
\ \ \ \ \sum_a Q^\nu_a =0 \ \ \ \( \hbox{and therefore}\ \ \ \nabla_\mu\sum_a
T^{\mu\nu}_a = 0\),
\label{coupled}
\end{equation}
assuming some action/reaction principle for the various fluid components.
For the background, setting $Q_a^\mu = \(-a Q_a, \bm{0}\)$, we have
the generalization of (\ref{dTeta}) to a many-component fluid, namely
\begin{equation}
\rho_a'+3\Hu\(1+w_a\) \rho_a = a Q_a \ \ \ \ \hbox{with} 
\ \ \ \ \sum_a Q_a =0.
\end{equation}
In practice however, since we shall here restrict attention to matter and
radiation, we set the forces acting on the fluids $Q^\nu_a\to 0$.

In terms of these variables, the relevant Einstein equation reads
\begin{equation}
\Delta \Phi = \frac32 \Hu^2 \sum_a\Omega_a \delta^\mathrm{_C}_a,
\label{PoissonPerturb}
\end{equation}
showing how to relate the large-scale structure distribution (the
density fluctuations) to the metric perturbations. The perturbed
densities and velocities, when both the forces $Q_a\to 0$ and the
self entropies $\Gamma_a\to 0$ are vanishing, follow the continuity
and Euler equations
\begin{equation}
\( \frac{\delta^\mathrm{_N}}{1+w_a}\)'+\Delta V_a -3\Phi' = 0 \ \ \ \ \hbox{and}
\ \ \ \ V_a'+\Hu V_a + \Phi+\frac{c_a^2}{1+w_a} \delta^\mathrm{_C}_a=0.
\label{deltaaPVaP}
\end{equation}

Let us specialize for now on to the case of two fluids.
Introducing the gauge-invariant
relative velocity $\tilde v$ and entropy perturbations $S$
\begin{equation}
\tilde v \equiv v_a - v_b \ \ \ \ \hbox{and} \ \ \ \ 
S \equiv \frac{\delta_a}{1+w_a} - \frac{\delta_b}{1+w_b},
\label{Sabvab}
\end{equation}
relation which can be inverted through
\begin{equation}
\(\frac{\Omega_b}{1+w_a}+ \frac{\Omega_a}{1+w_b}\) \delta_a
=\frac{\Omega \delta}{1+w_b} + \Omega_b S,
\end{equation}
the continuity equation can easily be restated as
\begin{equation}
S' = -\Delta \tilde v - 3 \Hu \tilde \Gamma, \ \ \ \hbox{where}
\ \ \ \ \tilde \Gamma\equiv \frac{w_a\Gamma_a}{1+w_a} - \frac{w_b\Gamma_b}{1+w_b},
\label{ContPert}
\end{equation}
while Euler equation reads
\begin{equation}
\tilde v' = -\Hu \tilde v -\(c_a^2-c_b^2\) \frac{\delta^\mathrm{_C}}{1+w} + \[ c_a^2 \(1+w_b\)
\frac{\Omega_b}{\Omega} + c_b^2 \(1+w_a\) \frac{\Omega_a}{\Omega} \] \frac{S}{1+w}
-\Gamma_{ab}.
\label{EulerPert}
\end{equation}
The basic idea now consists in solving Eqs.~(\ref{ContPert}) and (\ref{EulerPert})
together with (\ref{PhiEvol}) so as to get a complete solution for the distribution
of $\delta^\mathrm{_C}$ now through (\ref{PoissonPerturb}): this density perturbation
spectrum can then be directly observed as the large-scale structure distribution.
This is more easily said than done, and to begin with, one needs to impose initial
conditions, to which I now turn.

\subsection{Adiabatic and isocurvature initial conditions}

In order to impose a complete set of initial conditions, we need to know
the number of independent degrees of freedom. As soon as one knows
all the fluid variables, the system is fixed, namely knowledge of all the
$\delta_a$ and $v_a$ is enough. For the two constituent fluid we are dealing
with, this means we need 4 independent conditions. With the previous
variables, we can re-express all these in terms of the sums 
$\delta^\mathrm{_C}$ and $V$, so that then Eqs.~(\ref{PoissonPerturb})
and (\ref{DiPhi}) provide $\Phi$ and $\Phi'$ respectively. We are
then left with the relative velocity and entropy perturbations $\tilde v$
and $S$.

There are basically two sets of initial conditions which are used, the so-called
adiabatic and isocurvatures ones. They are defined by the following conditions.

\begin{itemize}

\item \underline{Adiabatic initial conditions:} the entropy perturbation (\ref{Sabvab})
vanishes at the initial time, while the Bardeen gravitational potential $\Phi$ is
a constant, so we have
\begin{equation}
S=0 \ \ \Longrightarrow \ \ \ \frac{\delta_a}{1+w_a} = \frac{\delta_b}{1+w_b}
\ \ \ \hbox{and} \ \ \ \Phi'=0,
\label{AdIC}
\end{equation}
leaving two arbitrary initial numbers, $\Phi$ and $S'$ say, to decide of
the forthcoming mode evolution.

\item \underline{Isocurvature initial conditions:} the opposite, and complementary,
situation consists in setting initial conditions such that there is no initial metric
perturbation, i.e. we set at the initial time 
\begin{equation}
\Phi = 0 \ \ \ \hbox{and} \ \ \ \delta^\mathrm{_C} = 0 \ \ \Longrightarrow \ \ \ 
\sum_a  \Omega_a \delta_a^\mathrm{_C} = 0,
\label{IsoCur}
\end{equation}
which, again, leaves 2 arbitrary numbers to be set, for instance the values
of $\Phi'$ and the initial entropy $S$.
\end{itemize}

These conditions essentially reflects all the possibilities, and the ``real''
initial perturbation should be a linear superposition of those.

It should be reminded at this stage that these initial conditions ought to be
set at the point in time after which we can evolve them with sufficiently
precise knowledge of the cosmic history. Normally, this is done using a
simulation code taking into account all known relevant cosmological
phenomena. This means we assume that the initial spectrum of perturbations
is propagated through the almost entire history of the Universe from this
original time... the question then remains of what is this initial time, and
how can we even suppose we know anything at all about it? That will
be the subject of the final section, but for now on, let us concentrate to
the actual evolution and the spectrum we might get now so as to be
able to compare with observational data!

\subsection{Mode history}

As discussed above, we now specifically restrict attention to a flat,
matter-dominated Universe containing a tiny amount of radiation, so
that Eqs.~(\ref{deltaaPVaP}) now read, in Fourier space,
\begin{align}
& \delta^{\mathrm{_N}\prime}\dm = k^2 V\dm+3\Phi' \ \ \ \ \hbox{and} \ \ \ V'\dm +\Hu V\dm+\Phi=0,
\label{deltam}\\
& \delta^{\mathrm{_N}\prime}\dr = \frac43 k^2 V\dr+4\Phi' \ \ \ \hbox{and} \ \ \ V'\dm +\Phi
+\frac14 \delta^\mathrm{_N}\dr=0,
\label{deltam}
\end{align}
and we assume we know, somehow, the initial conditions for the perturbations
deep into the radiation epoch. Adding the Fourier-transformed Poisson equation
\begin{equation}
-k^2 \Phi = \frac32 \Hu^2 \[ \Omega\dm\delta^\mathrm{_N}\dm +
\Omega\dr\delta^\mathrm{_N}\dr -3\Hu \( V\dm +\frac43 V\dr\)\]
\label{PoissonPhi}
\end{equation}
closes the system which we now solve.

\subsubsection{Initial conditions in the early radiation epoch}

The Universe will have to go through the
radiation-to-matter transition, and we thus switch to the relevant time variable
$y\equiv a/a_\mathrm{eq}$ defined above Eq.~(\ref{FriedTrans}) and in terms
of which the relative density parameters read
$$
\Omega\dm  = \frac{y}{1+y} \ \ \ \hbox{and} \ \ \ \ \Omega\dr=\frac{1}{1+y}.
$$
Since now $S=\delta\dm -\frac34 \delta\dr$ and noting the relationship
between the comoving total density perturbation and the Bardeen potential
$$
\delta^\mathrm{_C} = -\frac43 \( \frac{k}{k_\mathrm{eq}}\)^2 \frac{y^2}{1+y} \Phi,
$$
we see that the entire system reduces to the set
\begin{align}
\begin{cases}
&\hskip-3mm\displaystyle\frac{\dd^2\Phi}{\dd y^2} +\frac{1}{2y} \( 7-\frac{1}{1+y}+\frac{8}{4+3y}\)\frac{\dd\Phi}{\dd y} +
\frac{\Phi}{y(1+y)(4+3y)} = \frac{2}{y^2(4+3y)}\(\delta^\mathrm{_C}-\frac{y S}{1+y}\),\label{d2Phi}\\
&\\
&\hskip-3mm\displaystyle\frac{\dd^2 S}{\dd y^2} +\frac{3y+2}{2y(1+y)} \frac{\dd S}{\dd y}
= \frac{2}{4+3y}\(\frac{k}{k_\mathrm{eq}}\)^2\(\delta^\mathrm{_C}-\frac{y S}{1+y}\).
\end{cases}
\end{align}
Once we have the solution to the system (\ref{d2Phi}), we can reconstruct the density
perturbations through 
\begin{equation}
\delta^\mathrm{_N} = \delta^\mathrm{_C} -2 \(\Phi + y\frac{\dd \Phi}{\dd y}\) \ \ \ \hbox{and} \ \ \ \
\delta\dm = \frac{\frac34\(1+y\)\delta +S}{1+\frac34 y}, \ \ \ \ \delta\dr = \frac{\(1+y\)\delta -y S}{1+\frac34 y}.
\label{densities}
\end{equation}

\begin{figure}[h]
\centering
\includegraphics[width=6.0cm,clip]{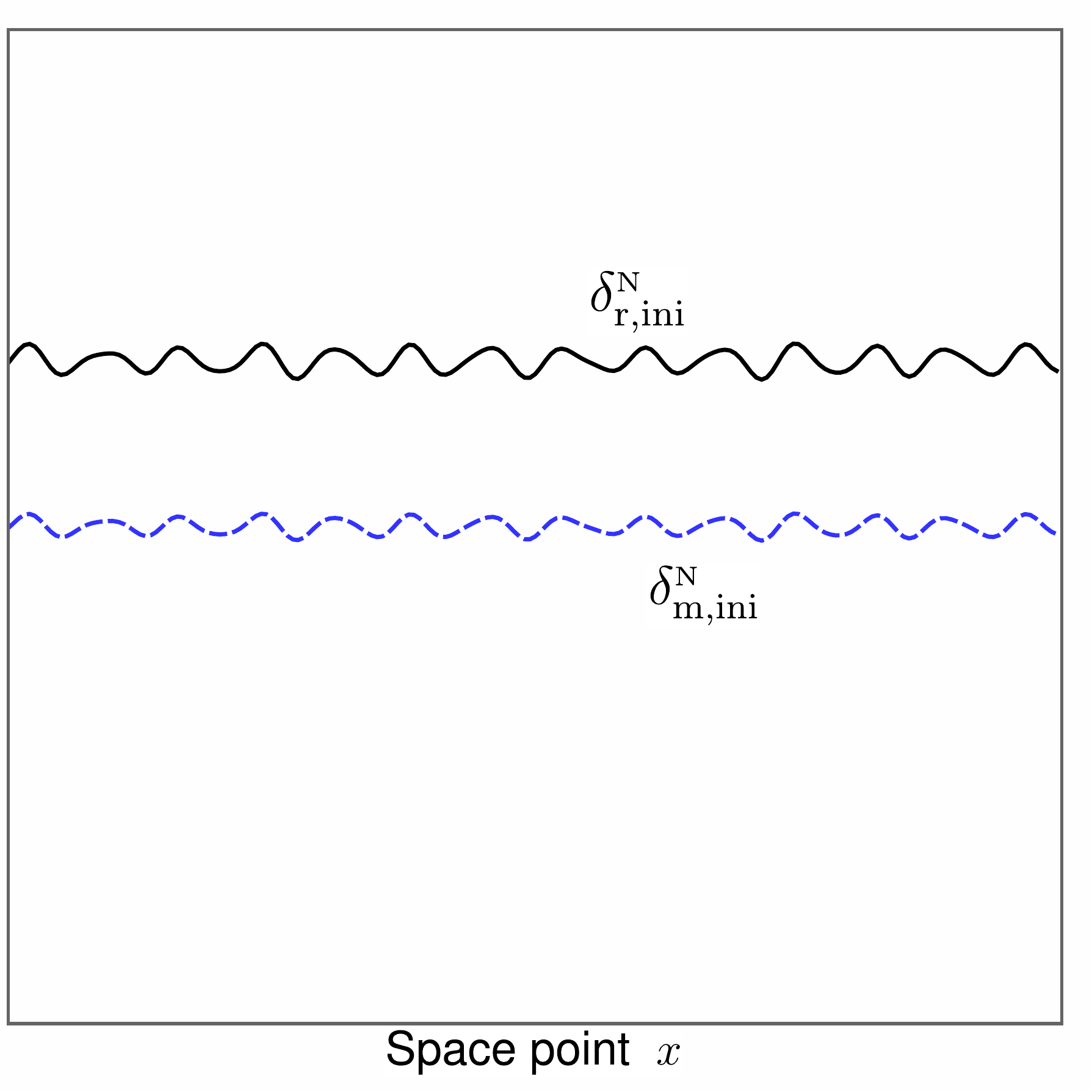}\hskip1cm
\includegraphics[width=6.0cm,clip]{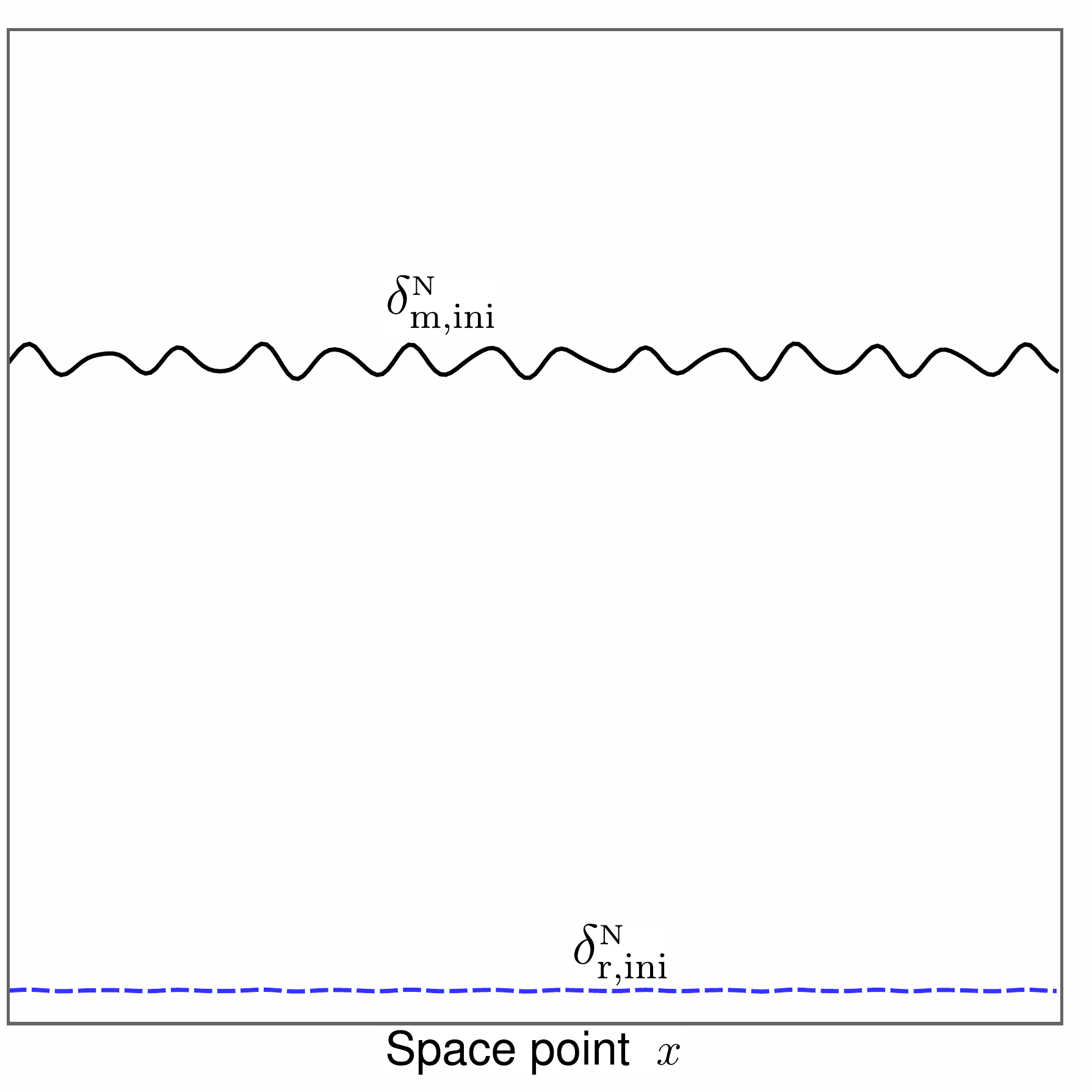}
\caption{Adiabatic and isocurvature initial conditions in terms of
the primordial density fluctuations: the left panel shows the adiabatic
condition, where the radiation (full line) and matter (dashed line)
density perturbations are everywhere following the same pattern of
fluctuations, while the isocurvature condition, represented on the
right panel, has a dominant contribution coming from the matter
fluctuation (full line) together with a negligible amount of radiation
density perturbation (dashed line).}
\label{AdiabIso}
\end{figure}

We now want to impose initial condition very early on, when the Universe is radiation 
dominated, i.e. for $y_\mathrm{ini}\ll 1$, and we are interested in cosmologically relevant
wavelengths, i.e. those satisfying $k\ll\Hu_\mathrm{ini}$. If we decide for adiabatic
initial conditions, this means we can demand
\begin{equation}
\Phi = \Phi_\mathrm{ini}, \ \ \ \hbox{and} \ \ \ \frac{\dd\Phi}{\dd y}\Big|_{y=y_\mathrm{ini}} =S_\mathrm{i}
=\frac{\dd S}{\dd y}\Big|_{y=y_\mathrm{ini}}=0,
\end{equation}
turning Eq.~(\ref{densities}) into
\begin{equation}
\delta^\mathrm{_C}_\mathrm{ini} = -\frac23 \(\frac{k}{\Hu_\mathrm{ini}}\)^2 \Phi_\mathrm{ini},
\ \ \ \ 
\delta^\mathrm{_C}_\mathrm{r,ini} = \delta^\mathrm{_C}_\mathrm{ini} \ \ \ \hbox{and}
\ \ \ \delta^\mathrm{_C}_\mathrm{m,ini} = \frac34\delta^\mathrm{_C}_\mathrm{ini},
\label{densitiesIniAdC}
\end{equation}
leading to
\begin{equation}
kV_\mathrm{ini} = -\frac12 \(\frac{k}{\Hu_\mathrm{ini}}\) \Phi_\mathrm{ini},
\ \ \ \ \delta^\mathrm{_N}_\mathrm{ini} =-2\Phi_\mathrm{ini}, \ \ \ \ 
\delta^\mathrm{_N}_\mathrm{r,ini} = \delta^\mathrm{_N}_\mathrm{ini} \ \ \ \hbox{and}
\ \ \ \delta^\mathrm{_N}_\mathrm{m,ini} = \frac34\delta^\mathrm{_N}_\mathrm{ini}.
\label{densitiesIniAdN}
\end{equation}
These initial conditions mean that the density ratios are constant everywhere on
the initial hypersurface: both density perturbations behave in the same way everywhere,
as illustrated in the left panel of Fig.~\ref{AdiabIso}.

Setting isocurvature initial conditions on the other hand amounts, in
this case, to imposing at the initial time the relations
\begin{equation}
S = S_\mathrm{ini}, \ \ \ \hbox{and} \ \ \ \frac{\dd S}{\dd y}\Big|_{y=y_\mathrm{ini}} =
\Phi_\mathrm{i} =\frac{\dd\Phi}{\dd y}\Big|_{y=y_\mathrm{ini}}=0,
\end{equation}
which translates, in terms of density perturbations, into
\begin{equation}
\delta^\mathrm{_C}_\mathrm{ini} = 0,
\ \ \ \ 
\delta^\mathrm{_C}_\mathrm{r,ini} = -yS_\mathrm{ini} \ll 
\delta^\mathrm{_C}_\mathrm{m,ini} = S_\mathrm{ini} \ \ \ \hbox{and}
\ \ \ \ \delta^\mathrm{_C}_\mathrm{ini} =\delta^\mathrm{_N}_\mathrm{ini}
\label{IsoCond}
\end{equation}
with all velocities vanishing. This is also illustrated in Fig.~\ref{AdiabIso}, on
the right panel.

\subsubsection{Transfer function}

Let me now move on to the transfer function, which is defined as the ratio
of the observed spectrum of perturbations now, i.e. essentially
the large scale structure distribution in the sky, with the primordial
spectrum, as calculated by high energy physics; this primordial
spectrum is the topic of the final section.

\begin{figure}[h]
\centering
\includegraphics[width=12.0cm,clip]{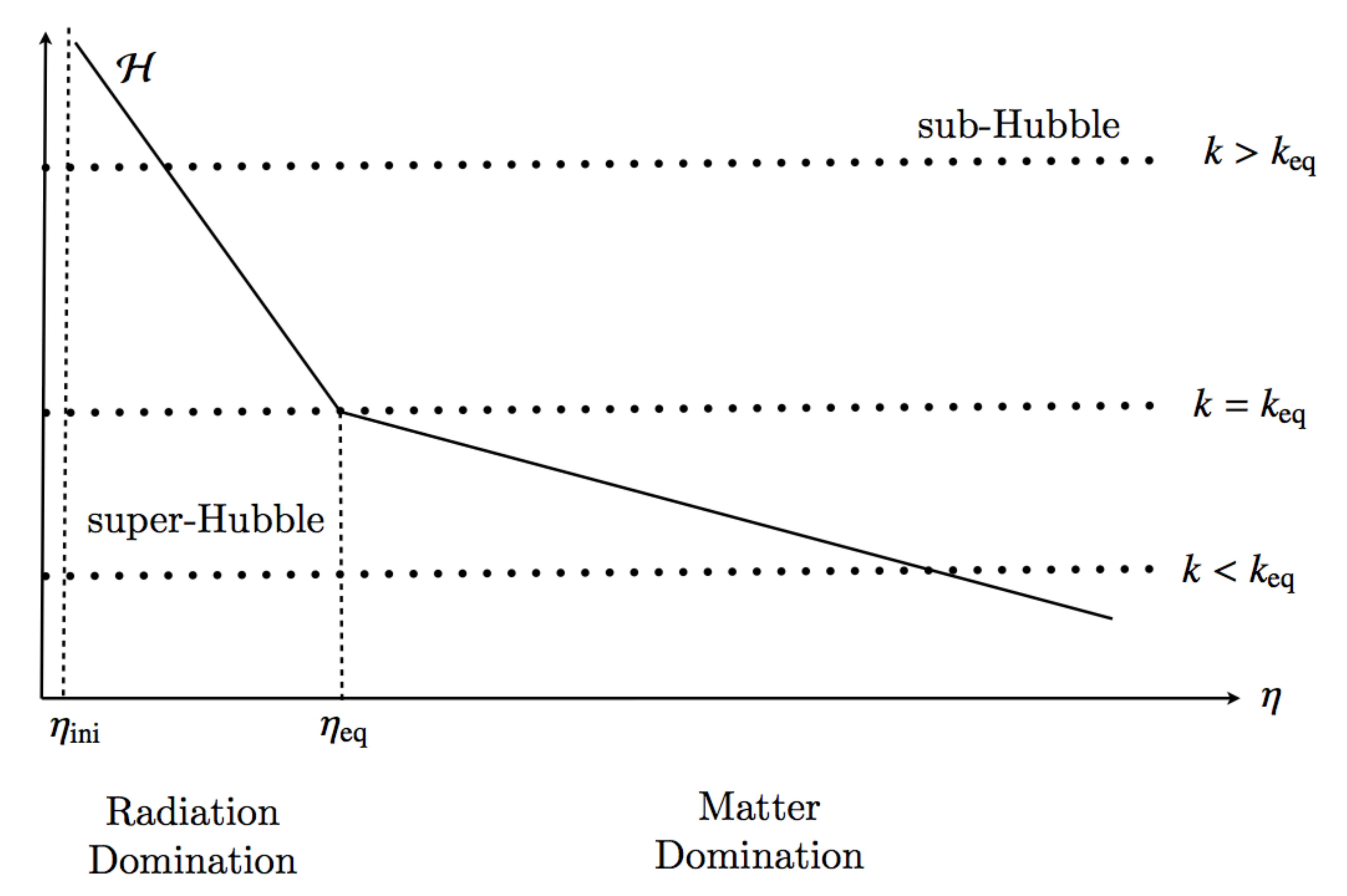}
\caption{Evolution regimes for perturbation modes. All modes
initiate at time $\eta_\mathrm{ini}$ during the radiation dominated
epoch, and are all super Hubble initially, i.e. they all satisfy
$k<\Hu_\mathrm{ini}$. Depending on whether the comoving
wavelength is larger or smaller than the Hubble scale at equality
[recall $k_\mathrm{eq} = \Hu_\mathrm{eq} =  \Hu(\eta_\mathrm{eq})$],
the mode becomes sub Hubble before or after the equality $\eta_\mathrm{eq}$.
This leads to different time evolutions, and a final spectrum that, more
or less independently of the initial spectrum, will have the equality
scale $k_\mathrm{eq}$ imprinted in it.}
\label{Modes}
\end{figure}

Let us consider a given mode, i.e. a given wavelength $k$, and
solve the equations of evolution (\ref{d2Phi}). Without entering
unnecessary details that can be found
elsewhere~\cite{PPJPU,Mukhanov:2005sc}, suffice it to say that
the main behavior of the perturbations depend on two quantities,
namely the time at which they are evaluated relative to the
equality time $\eta_\mathrm{eq}$, and the comoving
wavenumber of the perturbation, again relative to the value
$k_\mathrm{eq}$, defined above Eq.~(\ref{ModeTensTrans}).
Fig.~\ref{Modes} shows the various possibilities by comparing
the wavelength with the Hubble factor entering the evolution
equations.

\begin{figure}[h]
\centering
\includegraphics[width=12.0cm,clip]{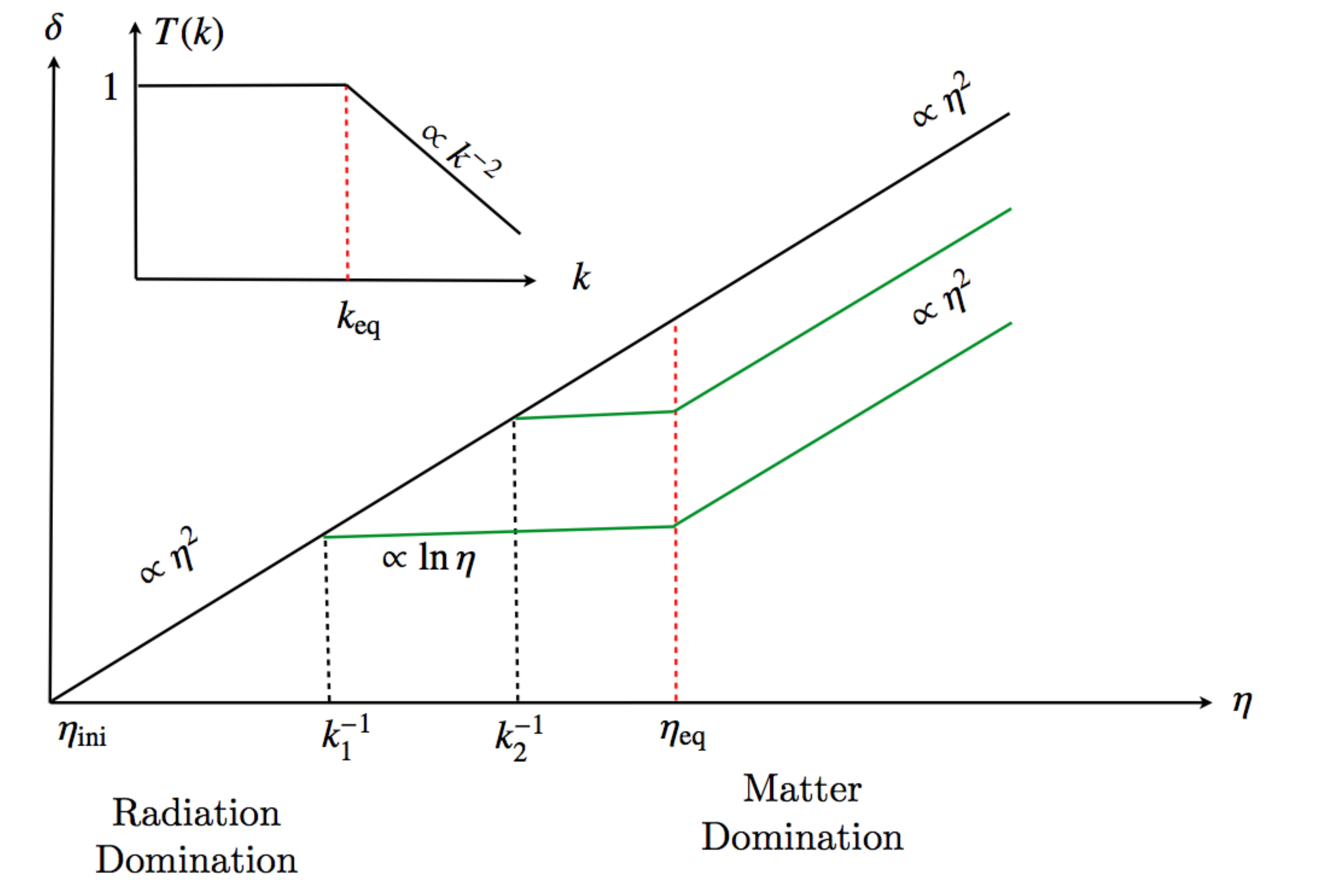}
\caption{Time evolution of a mode with the corresponding
transfer function as a function of the comoving wavenumber
$k$. The density fluctuation for modes becoming sub Hubble
during matter domination and having $k<k_\mathrm{eq}$
essentially grow at all times as $\propto\eta^2$, corresponding
in fact to different behaviors with the scale factor. For a finite
amount of time however, modes becoming sub Hubble during
radiation domination cannot grow before the advent of matter
domination and thus acquire a logarithmic, i.e. almost constant,
time behavior. As a result, their amplitude increases less than
the other mode's amplitude, leading to a transfer function $T(k)$
as indicated in the insert.}
\label{Transf}
\end{figure}

One finds the following time developments:
\begin{itemize}
\item \underline{$k\leq k_\mathrm{eq}$}
\begin{itemize}
\item $\eta_\mathrm{ini}\leq \eta \leq\eta_\mathrm{ini}$: $\Phi \sim \Phi_\mathrm{ini}\to
\delta\dm\propto\eta^2$,
\item $\eta_\mathrm{eq}\leq\eta \leq 1/k$: $\Phi \sim \frac{9}{10} \Phi_\mathrm{ini}\to
\delta\dm\propto\eta^2$,
\item $1/k\leq \eta\leq \eta_0$: $\delta\dm\propto\eta^2$.
\end{itemize}
\item \underline{$k\geq k_\mathrm{eq}$}
\begin{itemize}
\item $\eta_\mathrm{ini}\leq \eta \leq 1/k$: $\Phi \sim \Phi_\mathrm{ini}\to
\delta\dm\propto\eta^2$,
\item $1/k \leq\eta \leq \eta_\mathrm{eq}$: $\delta\dm\propto \ln a$,
\item $\eta_\mathrm{eq}\leq \eta\leq \eta_0$: $\delta\dm\propto\eta^2$.
\end{itemize}
\end{itemize}
We see that for most of the time, the density perturbation in the matter 
fluid evolves as the square of the conformal time, except for the modes
whose wavelength is smaller than the equality scale: those become
sub Hubble during the radiation dominated phase, during which they
cannot grow because of the photon pressure. This is shown in
Fig.~\ref{Transf}.

The previous time evolution transforms into a change in the spectrum
for scales above $k_\mathrm{eq}$. Indeed, a mode $\delta^<$
with $k<k_\mathrm{eq}$ evolves essentially as $\eta^2$ all along,
so that $\delta^<_0\ \sim \delta^<_\mathrm{ini} \eta_0^2$, with the
index ``0'' still denoting the present-day time. Similarly, a mode
$\delta^>$ with $k>k_\mathrm{eq}$ evolves as $\eta^2$ only up
until $\eta\sim k^{-1}$, at which point it behaves roughly as a
constant. As a result, when it starts growing again, at $\eta_\mathrm{eq}$,
its value is $\delta^>(\eta_\mathrm{eq})\sim \delta^>_\mathrm{ini} k^{-2}
\eta_0^2$. The transfer function $T(k)$ is now defined as the ratio
between the final (evaluated at $\eta_0$) and initial (at $\eta_\mathrm{ini}$)
density perturbations, namely
\begin{equation}
\delta \(k,\eta_0\) = T(k) \delta \(k, \eta_\mathrm{ini}\),
\label{Tk}
\end{equation}
and the calculation above shows that $T(k)\sim 1$ for $k<k_\mathrm{eq}$
(long wavelengths) and $T(k)\sim k^{-2}$ for $k>k_\mathrm{eq}$
(short wavelengths). The insert of Fig.~\ref{Transf} also shows the
typical behavior of the transfert function.

\subsubsection{Perturbation spectrum}

With the transfer function known, we can now derive the actual
large scale structure distribution, can we? Well, in fact not quite
yet, as there is something missing: the initial distribution 
$\delta \(k,\eta_\mathrm{ini}\)$. This will be the subject of the
last section, as I already mentioned a few times, but something
can already be said at this stage, in particular by looking at
the data.

The only analysis that can be done of all the observation is
statistical in nature, as we now understand the actual density
distribution to be but a particular realization of a statistical
ensemble, so that the density field itself is now seen as a
random variable at each point. What is actually measured
then is the correlation function $\xi\(\bm{r}\)$ of the density field,
defined by
\begin{equation}
\xi\(\bm{r}\) \equiv \langle \delta\(\bm{x}\) \delta\( \bm{x}+\bm{r}\)
\rangle,
\label{xi}
\end{equation}
where the mean value should represent an ensemble average. In practice
however, we have only access to one such realization, and we replace
the ensemble average by a spatial average. Note at this stage that
the cosmological principle implies that the distributions should be
isotropic and homogeneous. As a result, the correlation function should
only depend on the distance scale $r$ and neither on the particular direction
choice $\bm{r}/r$ nor on the specific point $\bm{x}$.

Moving to the Fourier space, we can write
\begin{equation}
\delta\(\bm{r}\) = \int \frac{\dd^3\bm{k}}{\(2\pi\)^{3/2}} \delta \( \bm{k} \)
\ex^{i\bm{k}\cdot \bm{r}},
\label{deltak}
\end{equation}
whose spectrum $\mathcal{P}_\delta \(k\)$ stems from the two-point function
in Fourier space, namely
\begin{equation}
\langle \delta \( \bm{k} \) \delta \( \bm{p} \) \rangle =
\mathcal{P}_\delta \(k\) \delta\( \bm{k}+\bm{p}\).
\label{xi}
\end{equation}
I leave as an exercise to show that it is the Fourier transform of
the correlation function, i.e.
\begin{equation}
\mathcal{P}_\delta \(k\) = \int \frac{\dd^3\bm{r}}{\(2\pi\)^{3/2}} \xi\(r\)
\ex^{i\bm{k}\cdot \bm{r}},
\label{Pdelta}
\end{equation}
and, as expected again from the cosmological principle, it also does
not depend on the direction $\bm{k}/k$ but merely on the wavenumber $k$.

In a way, the power spectrum can roughly be seen as the square of
the density distribution. Therefore, we also have the relation
\begin{equation}
\mathcal{P}_\delta \(k,\eta_0\) = T^2(k) \mathcal{P}_\delta \(k, \eta_\mathrm{ini}\),
\label{Tk2}
\end{equation}
which is the equivalent of (\ref{Tk}) for the spectra.

\begin{figure}[h]
\centering
\includegraphics[width=12.0cm,clip]{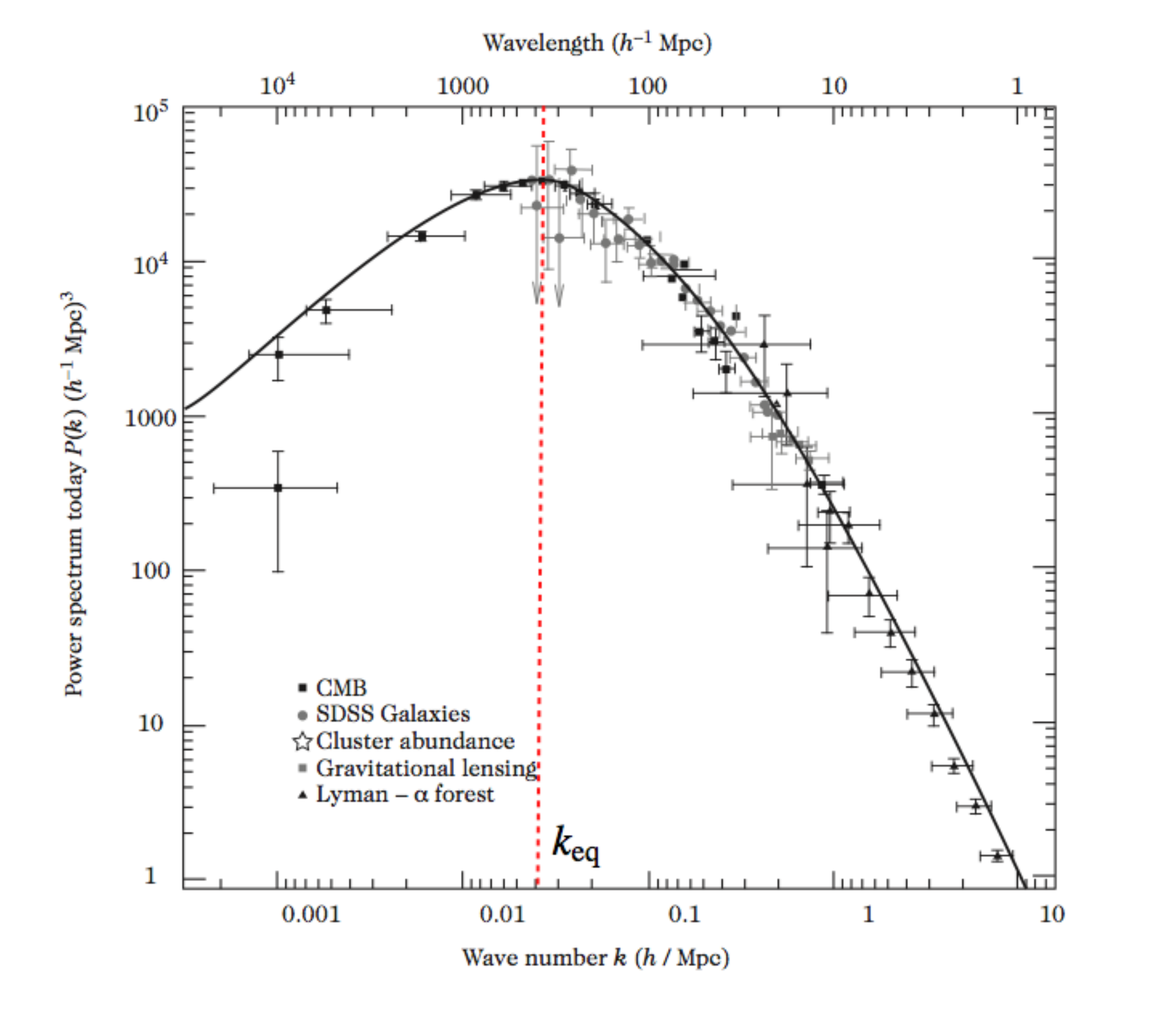}
\caption{Actual observations of the large scale structure distribution
showing a linear behavior in the wavenumber for large scales, followed
by a decrease as $k^{-3}$ for small scales. The spectrum peaks around
a few hundreds $h^{-1}$ Mpc, which thus indicates the value of the equality
scale, i.e. the Hubble radius at $\eta_\mathrm{eq}$.}
\label{Spect}
\end{figure}

We shall see later that the expected primordial spectrum actually scales
like $\mathcal{P}_\delta \(k, \eta_\mathrm{ini}\) \propto k$, and so the
observed distribution should scale as $k$ for long wavelengths where the
transfer function is independent of scale, and as $k^{-3}$ for shorter
wavelengths. Fig.~\ref{Spect} roughly confirms these expectations.

A quick estimate of the equality scale $k_\mathrm{eq}$ is provided
by
$$
k_\mathrm{eq} = H_0 a_0 \sqrt{2\Omega\dm^0\(1+z_\mathrm{eq}\)}
\sim 0.072 \Omega\dm^0 h^2 \hbox{Mpc}^{-1} \ \ \ \ \Longrightarrow\ \ \ \
\lambda_\mathrm{eq} \equiv k^{-1}_\mathrm{eq} \sim
\frac{14\,\hbox{Mpc}}{\Omega\dm^0 h^2},
$$
which is estimated to a hundred Mpc, falling a bit short of the actual
value. This is due to our very rough approximation according to which our universe
model only contains matter and radiation.

The actual transfer function is in fact much more complicated to calculate
once one takes into account all the cosmological constituents. For instance,
if there is a so-called hot dark matter component, consisting in relativistic
degrees of freedom at the time of decoupling, e.g. neutrinos, it has the effect
of streaming easily throughout any initial perturbation, thus wiping out very
large scales. These scales cannot grow anymore, and this produces an
exponential cutoff in the transfer function.

\section{Initial conditions: quantum vacuum fluctuations}

So far, this presentation contained essentially no new physics, merely basic
applications of general relativity and fluid dynamics. It can all be made much
more precise, e.g. to include thermodynamics and using the Boltzmann
equation, but this would also be well-known physics and in no way can ever
provide what we are seeking, namely the initial conditions for the perturbations
we have just calculated. In fact, measuring the large scale structure distribution
while knowing all what precedes is akin to measuring the initial conditions, which
is kind of useless if we don't have a theory that predicts them.

It turns out that soon after the advent of inflation, which was originally the
first convincing scenario that was proposed to cure the puzzles discussed
in the first section of these lectures, it was realized that the accelerating
epoch had the ability not only to enhance already-existing perturbations,
but also to produce those when quantum vacuum fluctuations were taken
into account. Since that time, other models, for instance including an
initial contracting phase and a bounce to connect to our currently expanding
epoch, have been devised that also provide the required initial conditions,
and they can be tested quite accurately with the more and more precise
data that are accumulating.

Although the inflationary paradigm is the best accepted one to describe the
primordial epoch, I would like first to emphasize that it is not established beyond
any doubt (as one sometimes reads!), so that looking for challengers is still
a reasonable activity. However, in what follows, I will restrict attention
to the inflationary case as it is easier to implement and pedagogically more
convenient.

\subsection{Back to the background}

Inflation provides explanations to the standard model puzzles by means
of acceleration of the scale factor, namely for a finite but sufficiently long
period of time, we have $\ddot a>0$. Eq.~(\ref{att}) then implies that
the pressure should be more negative than a third of the energy density
(which is always assumed positive). This is easily achieved by means
of a slowly-rolling scalar field $\varphi$ whose action we take to be
\begin{equation}
S=\int \[\frac12 \(\partial\varphi\)^2 + V\(\varphi\) \]
\sqrt{-g} \dd^4 x,
\label{ActPhi}
\end{equation}
with a yet-undefined potential $V\(\varphi\)$.

\subsubsection{Slow-roll parameters}

The stress-energy tensor derivable from the action (\ref{ActPhi}) is
\begin{equation}
T_{\mu\nu} = \partial_\mu\varphi \partial_\nu\varphi -
\[\frac12 \(\partial\varphi\)^2 + V\(\varphi\) \]g_{\mu\nu}\ \ \ \ \Longrightarrow
\ \ \ \rho=\frac12 \dot\varphi^2+V \ \ \hbox{and} \ \ \ p=\frac12 \dot\varphi^2
-V,
\label{TmunuPhi}
\end{equation}
where the definition of the energy density and pressure assume the
field $\varphi$ to depend only on time in order to satisfy the background
symmetries. We see that the r.h.s. of Eq.~(\ref{att}) reads $\rho+3p =
2\(\dot\varphi^2 -V\)$, which can be negative quite easily provided the
kinetic term $\dot\varphi^2$ is sufficiently small compared with the
potential. Because then the velocity of the field is tiny, this is why we
speak of slow-roll phase.

The Einstein and Klein-Gordon equations then transform into
\begin{align}
\begin{cases}
&\hskip-3mm\displaystyle H^2 = \frac{8\pi\GN}{3}\[\frac12\dot\varphi^2
+V\(\varphi\)\] - \frac{\Ka}{a^2},\\
&\hskip-3mm\displaystyle \frac{\ddot a}{a} = \frac{8\pi\GN}{3}\[ V\(\varphi\)
-\dot\varphi^2\],\\
&\hskip-3mm\displaystyle \ddot \varphi + 3 H \dot\varphi +\frac{\dd V}{\dd \varphi}
=0,
\end{cases}
\end{align}
where the last of these, merely reflecting the conservation of (\ref{TmunuPhi}),
is not independent of the first two. Combining those actually yields $\dot H = -4\pi
\GN\dot\varphi^2 +\Ka/a^2$: as $\dot\varphi^2$ is assumed small, the natural
tendency for $\dot H$ is to decrease as the scale factor increases. But this makes
the scale factor increase even more rapidly, so the spatial curvature term
becomes more and more negligible. In fact, this is an attractor of this system
of equations, and therefore, for now on, we will assume $\Ka\to 0$ with the
meaning that spatial curvature terms are exponentially smaller than any other.

Applying the slow-roll conditions $\dot\varphi^2\ll V$ and $\ddot\varphi \ll
3H\dot\varphi$, we find the relations
\begin{equation}
H^2\simeq \frac{8\pi\GN}{3}V, \ \ \ \ \dot H\simeq -4\pi\GN \dot\varphi^2 \ \ \ \ 
\hbox{and} \ \ \ \ 3H\dot\varphi \simeq V_{,\varphi}
\label{SReqs}
\end{equation}
which are consistent with the original assumptions only if $|\dot H|/H^2 \ll
3/2$. More generally, one can define two small parameters, called the
slow-roll parameters, by
\begin{equation}
\varepsilon \equiv -\frac{\dot H}{H^2} = 
\frac{\frac32\dot\varphi^2}{\frac12\dot\varphi^2 +V(\varphi)} \ \ \ \ \hbox{and}
\ \ \ \ \delta \equiv \varepsilon -\frac{\dot\varepsilon}{2H\varepsilon} = 
\frac{\ddot\varphi}{H\dot\varphi}.
\label{srp}
\end{equation}
Inflation goes on for as long as $\ddot a>0$, which translates into $\varepsilon <1$.

The simplest solution for this model consists in demanding $\varepsilon$ to be
constant; in this case, the scale factor can be calculated as follows.
First, I recall the relationship between conformal and cosmic time, namely
\begin{equation}
\dd t = a \dd \eta \ \ \ \Longrightarrow \ \ \ H=\frac{1}{a}\frac{\dd a}{\dd t} = 
\frac{\dd a}{a^2 \dd t/a} = \frac{\dd a}{a^2\dd\eta} \ \ \ \Longrightarrow \ \ \ 
\eta=\int\frac{\dd a}{a^2 H},
\end{equation}
so that, noting we also have
\begin{equation}
\dd\(\frac{-1}{a H} \) = \frac{1}{H} \frac{\dd a}{a^2} +\frac{1}{a}\frac{\dd H}{H^2}
= \frac{\dd a}{a^2 H} + \frac{1}{a H^2}\frac{\dd H}{\dd t}\dd t = \frac{\dd a}{a^2 H}
-\frac{\varepsilon}{a}\dd t = \frac{\dd a}{a^2 H}-\frac{\varepsilon}{a}\frac{\dd t}{\dd a}
\dd a,
\end{equation}
this means that the relation
$$
\eta=\int\dd\(\frac{-1}{aH}\)+\int\frac{\varepsilon}{a^2 H}\dd a
$$
should hold, and if $\varepsilon$ is roughly constant, we can perform the
integration, leading to
$$
\eta = -\frac{1}{a H}+\varepsilon \int\frac{\dd a}{a^2 H} = -\frac{1}{aH} + \varepsilon
\eta,
$$
which I can invert and obtain that the scale factor behaves as
\begin{equation}
a = \frac{-1}{H\eta\(1-\varepsilon\)} = \frac{-1}{H\eta_0 \(1-\varepsilon\)}
\ex^{H \(1-\varepsilon\)\(t-t_0\)},
\label{QuasiDS}
\end{equation}
i.e. we have an exponential quasi de Sitter phase. The
parameters $t_0$ and $\eta_0$ are constant of integration
necessary to pass from conformal to cosmic time.
Eq.~(\ref{QuasiDS}) shows moreover that inflation
occurs in the regime where $\eta<0$ and $\eta\to 0^-$.

\subsubsection{Two explicit examples}

The simplest example one can think of is that for which the scalar
field is merely a massive free (non interacting) field, namely the
potential reads $V(\varphi) = \frac12 m^2 \varphi^2$. In this case,
the system (\ref{SReqs}) reads
\begin{equation}
3H \dot\varphi + m^2\varphi = 0 \ \ \ \Longrightarrow \ \ \ \varphi\(t\)
=\varphi_\mathrm{ini} - \frac{m \Mp}{\sqrt{12\pi}} t,
\end{equation}
where use has been made of 
$$
H^2 = \frac43 \pi \(\frac{m}{\Mp}\)^2 \varphi^2.
$$
The scale factor is then
$$
a\(t\) = a_\mathrm{ini} \exp\left\{ \frac{2\pi}{\Mp^2}\[ \varphi_\mathrm{ini}^2
-\varphi^2\(t\)\]\right\},
$$
from which one obtain the slow-roll parameters as
\begin{equation}
\varepsilon = \frac{\Mp^2}{4\pi\varphi^2} \ \ \ \ \text{and} \ \ \ \
\delta=0.
\end{equation}
As $\varepsilon$ varies with time, we can easily calculate when the slow-roll
phase ends, namely for $\varphi = \varphi_\mathrm{f} = \Mp/\sqrt{4\pi}$, so
the number of e-folds of inflation is $N=2\pi\(\varphi_\mathrm{ini}/\Mp\)^2
-\frac12$. In order to solve the cosmological puzzles, we know that we
must impose $N\gsim 70$, leading to the requirement that the initial
value of the scalar field should be of order $\varphi_\mathrm{ini}\simeq
3\Mp$. One might think that this could be a problem, as such a high energy
scale would require quantum gravity to be described properly, and of course
we do not have such a theory. However, what actually matters is not the
field value itself, but the energy density
that it stores. Under the slow-roll hypothesis, this means the potential
energy, i.e. $V_\mathrm{ini}\sim \frac92 \(m\Mp\)^2 \ll \Mp^4$ provided the
scalar field mass is much less than the Planck scale. As we shall see
below, this is exactly what is required from the data.

Another useful model is the so-called power-law inflation, for
which one demands the scale factor to increase as a power-law
instead of an exponential, while still being accelerated. Explicitly,
this is
\begin{equation}
a=a_\eta \(-\eta\)^{1+\beta} \ \ \ \Longleftrightarrow \ \ \ a=a_t t^p \ \ \ \
\text{with} \ \ \ \ 1+\beta = \frac{p}{1-p},
\end{equation}
where $a_\eta$ and $a_t$ are constants, and $p>1$ to ensure
that $\ddot a>0$.

Integrating (\ref{SReqs}) equations again, one obtains the scalar field
behavior
$$
\frac{\varphi-\varphi_\mathrm{ini}}{\Mp} = \frac{1+\beta}{2\sqrt{p \pi}}
\ln\(-\eta\),
$$
and the potential it evolves in
$$
V\(\varphi\) = V_\mathrm{ini} \exp\[ 4\sqrt{\frac{\pi}{p}}
\(\frac{\varphi-\varphi_\mathrm{ini}}{\Mp}\)\],
$$
together with the slow-roll parameters: $\varepsilon = \delta = 1/p$. Since
there is no time evolution in this case, we see that such a model has
merely a pedagogical value, as inflation never ends in this case. However,
it can really be useful because many features are calculable in an
analytic way.

Having settled and somehow implemented the inflationary phase,
let us see what happens to fields living in such a background.

\subsection{A test field in inflationary background}

We shall here follow the evolution of the simplest case, namely that of
a test field in an inflationary background which, to make things even simpler,
we shall assume takes the form of a quasi de Sitter expansion, i.e. an
actually exponential expansion, or Eq.~(\ref{QuasiDS}).

\subsubsection{Massless scalar field}

Let us begin with yet another simplifying assumption, namely that the
test scalar field $\chi$ is massless, so its potential vanishes. The
Klein-Gordon in the expanding background reads
\begin{equation}
\ddot\chi+3H \dot\chi-\frac{1}{a^2}\Delta \chi = 0 \ \ \ \ \Longrightarrow \ \ \ \
\chi''+2\Hu \chi'+k^2\chi=0,
\label{KGtest}
\end{equation}
where in the last equality I switched to the conformal time and took
the Fourier transform. Now, setting $v=a\chi$, thereby defining $v$, we
find that 
\begin{equation}
v''+\[k^2-\(\Hu'+\Hu^2\)\] v = 0 \ \ \ \ \text{or} \ \ \ \ 
v''+\(k^2 -\frac{a''}{a}\)v= v''+\( k^2-\frac{2+3\varepsilon}{\eta^2}\)v=0.
\label{modesv}
\end{equation}
where I have used Eq.~(\ref{QuasiDS}).

Eq.~(\ref{modesv}) is the same as that driving the evolution of the
tensor modes (\ref{muT}), and is of the generic form of a "time-independent"
Schr\"odinger equation in a potential (regarding the conformal time
variable as the equivalent of the spatial coordinate):
\begin{equation}
-\frac{\hbar^2}{2m}\frac{\dd^2\psi}{\dd x^2} + \[V\(x\)-E\]\psi=0 \ \ \ \
\Longleftrightarrow \ \ \ \ \frac{\dd^2 v}{\dd\eta^2} + \[k^2 - U\(\eta\)\]v=0,
\label{Sch}
\end{equation}
provided one identifies $V(x)$ with $U(\eta)$, $E$ with $k^2$ and
rescale everything to cancel out the $-\hbar^2/(2m)$. In fact, whenever
there is no entropy perturbation, it is the most generic form we will
ever encounter. So for now on, I will assume a generic potential $U(\eta)$,
and discuss the actual mode evolution.

\subsubsection{Evolution regimes}

\begin{figure}[h]
\centering
\includegraphics[width=12.0cm,clip]{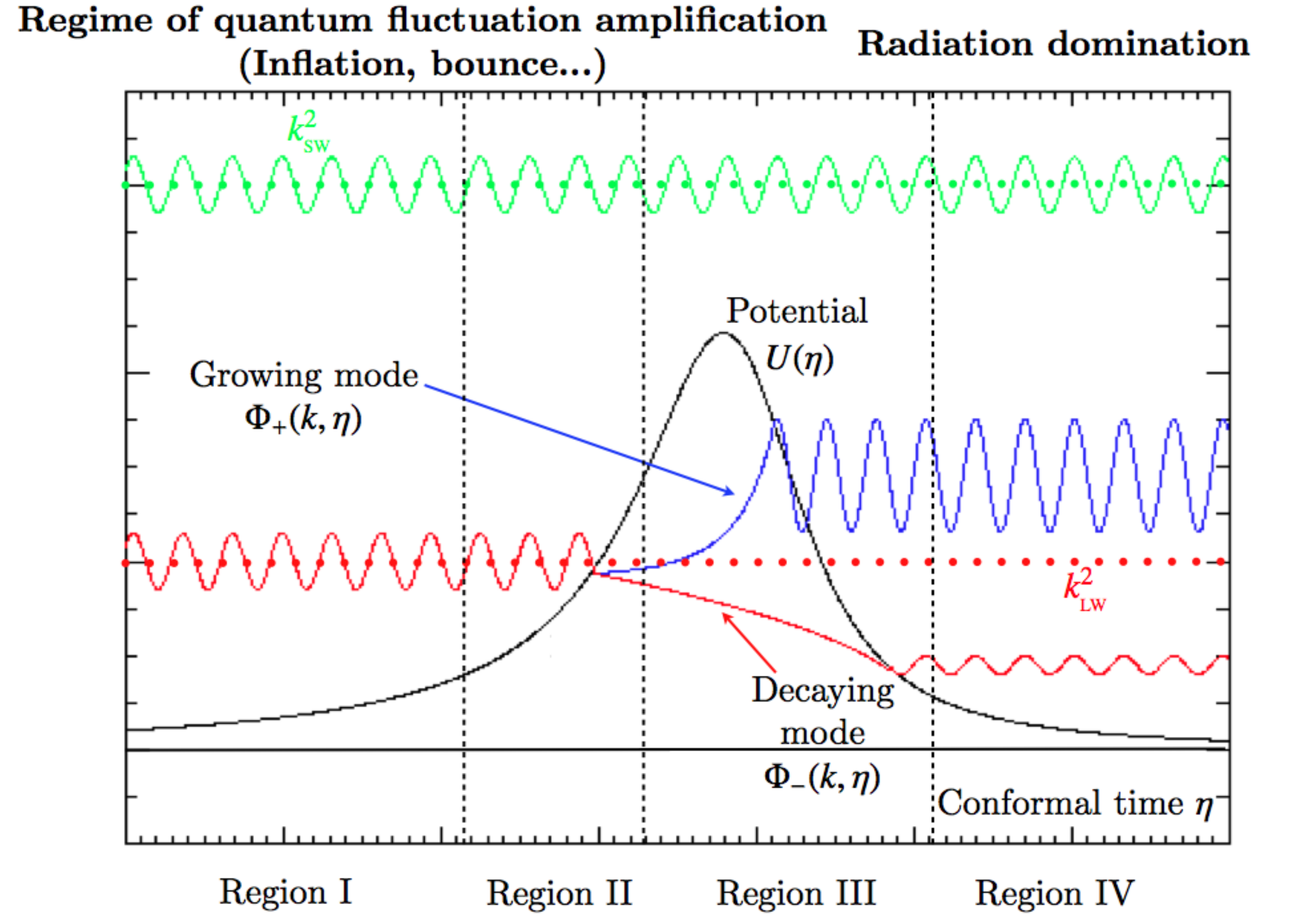}
\caption{Different regimes for the time evolution of a scalar mode:
the potential in Eq.~(\ref{Sch}) starts growing initially during the
perturbation production epoch, then stabilizes for instance at the
end of inflation or at a bouncing point, and then decays again while
getting into the radiation or matter domination era (or any other
relevant subsequent regime). For a short wavelength $\lambda_\mathrm{_{SW}}$,
the wavenumber $k^2$ is at all time larger than the potential, which therefore
doesn't affect the mode evolution: Eq.~(\ref{Sch}) then indicates a simple
oscillating behavior at all times. On the other hand, for a larger wavelength, 
$\lambda_\mathrm{_{LW}}$, i.e. smaller $k^2$, different regimes can be
identified: initially, in region I, the mode oscillates as $k^2\gg U(\eta)$, then
there is a transition through region II in which the mode passes below the
potential. Then in region III, one is in the opposite situation where $k^2\ll U(\eta)$,
and the mode now consists in a growing and a decaying solution. Finally,
region IV connects to the standard cosmology, the mode is above the
potential again, and therefore starts oscillating again; these oscillations
are those one observes in the Cosmic Microwave Background which I did
not have space to discuss here.}
\label{PotPert}
\end{figure}

The potential during inflation grows like $\eta^{-2}$ when $\eta\to 0^-$, but
this is merely an artifact of our approximation. In a realistic scenario however,
the potential might look like that represented in Fig.~\ref{PotPert}: it starts
growing during the phase of inflation (or in general any such phase
during which primordial perturbations are produced), and reaches a maximum,
after which it decays. These last phases would usually represent radiation
or matter domination, at which times we would observe the mode somehow:
the scale factor would then behave as $a\propto \eta^\beta$, so the typical 
potential should look like $a''/a = \beta\(\beta-1\) \eta^{-2}$.

Figure~\ref{PotPert} summarizes the discussion of the tensor modes, with
the same kind of solutions (\ref{short}) and (\ref{long}); I shall not repeat
this analysis here, but suffice it to say that it also applies to most known
cases as very often the potential has the form of the second time derivative
of a function over this function.

In the special case of de Sitter expansion, i.e. (\ref{QuasiDS}) with $\varepsilon
\to 0$, the solution is known, since this is then a quite simple Bessel equation,
and we have
\begin{equation}
v_k\(\eta\) = A(k)\ex^{-ik\eta} \( 1+\frac{1}{ik\eta}\) + B(k)\ex^{ik\eta}
\(1-\frac{1}{ik\eta}\),
\label{dSv}
\end{equation}
where $A(k)$ and $B(k)$ are yet-unknown function depending only
on the scale $k$.
 
The massive scalar field case can be obtained in a very similar
way as it suffices to replace (\ref{Sch}) by
$$
v'' + \( k^2 +\frac{m^2/H^2 -2}{\eta^2}\)v=0,
$$
whose solution is again another linear superposition of Bessel
functions of index $\nu$, with $\nu^2=\frac94-m^2/H^2$.

\subsection{Quantization}

All what precedes does still not tell us what initial conditions we
should use, or, in other words, given (\ref{dSv}), what should we
take as functions $A(k)$ and $B(k)$?

To achieve this goal, we need to quantize our system, which is
quite simply done when we have discussed the action expanded
to second order.

\subsubsection{Expanding the action}

The action for our scalar field, still without a potential to keep
things simple, is
\begin{equation}
S=\int \frac12 \(\partial\chi\)^2\sqrt{-g}\dd^4 x = \frac12 \int a^4\[
-\chi'^2+\(\bm{\nabla}\chi\)^2\]\dd^4 x,
\end{equation}
which we can express in terms of the variable $v$ as
$$
S=\underbrace{\frac12\int \dd^4 x\,\[-v'^2+\(\bm{\nabla}v\)^2
-\frac{a''}{a}v^2\]}_{\text{variable mass scalar field in Minkowski space}}+\ \
\underbrace{\frac12 \int \dd^4x \frac{\dd}{\dd\eta}\( \Hu v^2\)}_{\text{surface
term, irrelevant}},
$$
showing it is nothing but the action for a simple scalar field in
Minkowski space... usual technique of quantum field theory can
now be applied, and we will have the possibility of choosing a
specific quantum state to provide the initial conditions.

\subsubsection{Canonical quantization}

We can expand the field $v$ as any standard quantum field
through
\begin{equation}
v\(\bm{x},\eta\) = \int \frac{\dd^3\bm{k}}{\(2\pi\)^{3/2}} \[
v_{\bm{k}}\(\eta\) \ex^{i \bm{k}\cdot\bm{x}} a_{\bm{k}}
+ v^\star_{\bm{k}}\(\eta\) \ex^{-i \bm{k}\cdot\bm{x}} a^\dagger_{\bm{k}}\],
\end{equation}
the second term being the hermitian conjugate of the first.

Quantization is achieved by promoting $a_{\bm{k}}\to
\hat a_{\bm{k}}$ to an operator
in the Fock space of field configurations and imposing the canonical
commutation relations
\begin{equation}
\[\hat a_{\bm{k}},\hat a^\dagger_{\bm{q}}\] = \delta^{(3)}\(\bm{k}-\bm{q}\).
\label{aa}
\end{equation}
These relations can be seen as stemming from the actual field
quantization: defining the conjugate momentum
$$
\pi = \frac{\delta\mathcal{L}}{\delta v'} = v' \ \ \ \to \ \ \ \text{operator} \ \ \hat \pi,
$$
with the Lagrangian being the integrand in the definition of the action,
the Hamiltonian follows
$$
H = \int\( v'\pi-\mathcal{L}\) = \frac12 \int \(\pi^2 +\partial_i v\partial^i v
-\frac{a''}{a}\)\dd^4 x,
$$
and we can impose the standard equal time
commutation relations for the field operators, namely
\begin{equation}
\[\hat v\(\bm{x},\eta\),\hat v\(\bm{y},\eta\)\] = 0=
\[\hat \pi\(\bm{x},\eta\),\hat \pi\(\bm{y},\eta\)\] \ \ \ \text{and} \ \ \ \
\[\hat v\(\bm{x},\eta\),\hat \pi\(\bm{y},\eta\)\] = i\delta^{(3)}\(\bm{x}-\bm{y}\).
\label{commFields}
\end{equation}
These commutation rules are consistent with those above (\ref{aa})
only provided the Wronskian $W(k)
=v_k v'^\star_k-v^\star_k v'_k$ is normalized
to $W=i$ since one gets directly from the field expansion
$$
\[\hat v\(\bm{x},\eta\),\hat \pi\(\bm{y},\eta\)\] = \int \frac{\dd^3\bm{k}}{\(2\pi\)^3} 
\ex^{i\bm{k}\cdot\(\bm{x}-\bm{y}\)}W(k).
$$
We are almost done, having merely to define the relevant state
to assume as initial condition.

\subsubsection{The vacuum state}

In quantum field theory and therefore here as well, the vacuum state
is that which is annihilated by all the so-called ``creation'' operators
$a_k$, namely
$$
\hat a_{\bm{k}}|0\rangle = 0 \ \ \ \ \text{for all} \ \bm{k},
$$
and all other states are obtained by repeated application of the
operators $\hat a^\dagger_{\bm{k}}$ on $|0\rangle$.

In the limit $|k\eta|\gg 1$, i.e. for large negative conformal times
where we indeed want to impose our initial conditions, we are back
to the usual massless scalar field in a Minkowski space time, and
we know that the vacuum state must therefore satisfy
$$
v_{\bm{k}} \underset{|k\eta|\to\infty}{\longrightarrow} \frac{\ex^{-ik\eta}}{\sqrt{2k}},
$$
as indicated in any standard textbook on quantum field theory.
Given the previously obtained solution, this leads to the so-called
Bunch-Davies vacuum state
\begin{equation}
\chi_k\(\eta\) = \frac{H\eta}{\sqrt{2k}}\(1+\frac{1}{ik\eta}\)\ex^{-ik\eta},
\label{BD}
\end{equation}
which now provides a closed form initial solution for our perturbation. It is
with such initial condition that one finally gets the scale-invariant spectrum
which one compares with the observational data (and it works!).

The power spectrum is now obtained from the 2-point correlation function
$\xi_v \(\bm{x}-\bm{y}\) \equiv\langle 0|\hat v\(\bm{x},\eta\) 
v\(\bm{y},\eta\)|0\rangle$, which
gives
\begin{equation}
\xi_v = \int\frac{\dd^3\bm{k}}{\(2\pi\)^3}|v_k|^2 \ex^{i\bm{k}\cdot\(\bm{x}-\bm{y}\)}
=\int\frac{\dd k}{k} \frac{k^3}{2\pi^2} |v_k|^2 \frac{\sin kr}{kr},
\end{equation}
after integration over the angles and setting $r=|\bm{x}-\bm{y}|$.

It turns out that for large scales, i.e. super-Hubble modes, one finds that the
properties of a quantum field are the same as that of a classical stochastic field
with gaussian statistics. In particular, this means we can replace the
quantum averages by statistical ensemble averages. For the stochastic
variables, we find a power spectrum that scales as
\begin{equation}
P_\chi\(k\) = \frac{2\pi^2}{k^3} \mathcal{P}_\chi\(k\) = \frac{|v_k|^2}{a^2}
= \(\frac{H}{2\pi}\)^2,
\end{equation}
in other words a scale-invariant spectrum.

\subsection{Realistic perturbations}

If one wants to take into account all the actual complication of what is really
going on, one needs to consider all scalar, vector and tensor modes of the
metric and treat them including all possible effects. Although this is a very
complicated task, some situations allow to say something however. For
instance, assuming the scalar field $\varphi$ to drive the early history of
the Universe, one finds that the gauge-invariant degree of freedom
generated by the variable
\begin{equation}
\delta\varphi - \varphi'\frac{C}{\Hu} \equiv \frac{v}{a},
\label{MS}
\end{equation}
thus defining the so-called Mukhanov-Sasaki variable $v$, is enough
to describe all the fluctuations in a single field inflation. Expanding the action
to second order in perturbation and getting rid of the surface terms just
as above, one arrives at
\begin{equation}
\delta^{(2)} S = -\frac12 \int\dd^4x\, \[v'^2 -\(\bm{\nabla}v\)^2 + \frac{z''}{z}v^2 \] +
\text{surface terms,}
\ \ \ \ \text{with} \ \ \ z\equiv \frac{a\varphi'}{\Hu},
\end{equation}
thus showing what I previously said, that the typical equation of motion is
always of the same form. As the same analysis applies, one thus obtains
a way to set up initial conditions by assuming quantum vacuum in the
early stage of the Universe. This leads to a natural way to obtaining a
scale-invariant spectrum that fits extremely well all the known data.

\section{Conclusion: Conditions for alternative scenarios, the bouncing model}

The cosmological scenario, as discussed in the notes above, represents a
major achievement in physics, performed in less than 100 years! In that time,
it has been established that the Universe itself could be treated, studied and
understood as a regular physical system, despite the fact that it seems to
contradict, by its very uniqueness, the usual assumptions of the scientific
method. In fact, cosmology somehow extended the inductive method, replacing
for instance repetition of experiments by repetition of measurements in different
directions, in other words, replacing ensemble averages by ergodicity.

To summarize, we now have a rather clear view, basically, of what happened
during the last 13.7 billion years, with detailed calculations comparing amazingly
well with observations. Although I did not discuss them all, but these observations
range from consequences stemming directly from nuclear physics (nucleosynthesis),
thermodynamics, fluids mechanics, gravitational phenomena, and, as sketched
in the last section above, the relationship between gravity and quantum physics!
That only a bunch of "unpleasant" features are present in the data with the
overall picture being generally consistent is absolutely astounding and very
often not given enough emphasis.

\begin{figure}[h]
\centering
\includegraphics[width=12.0cm,clip]{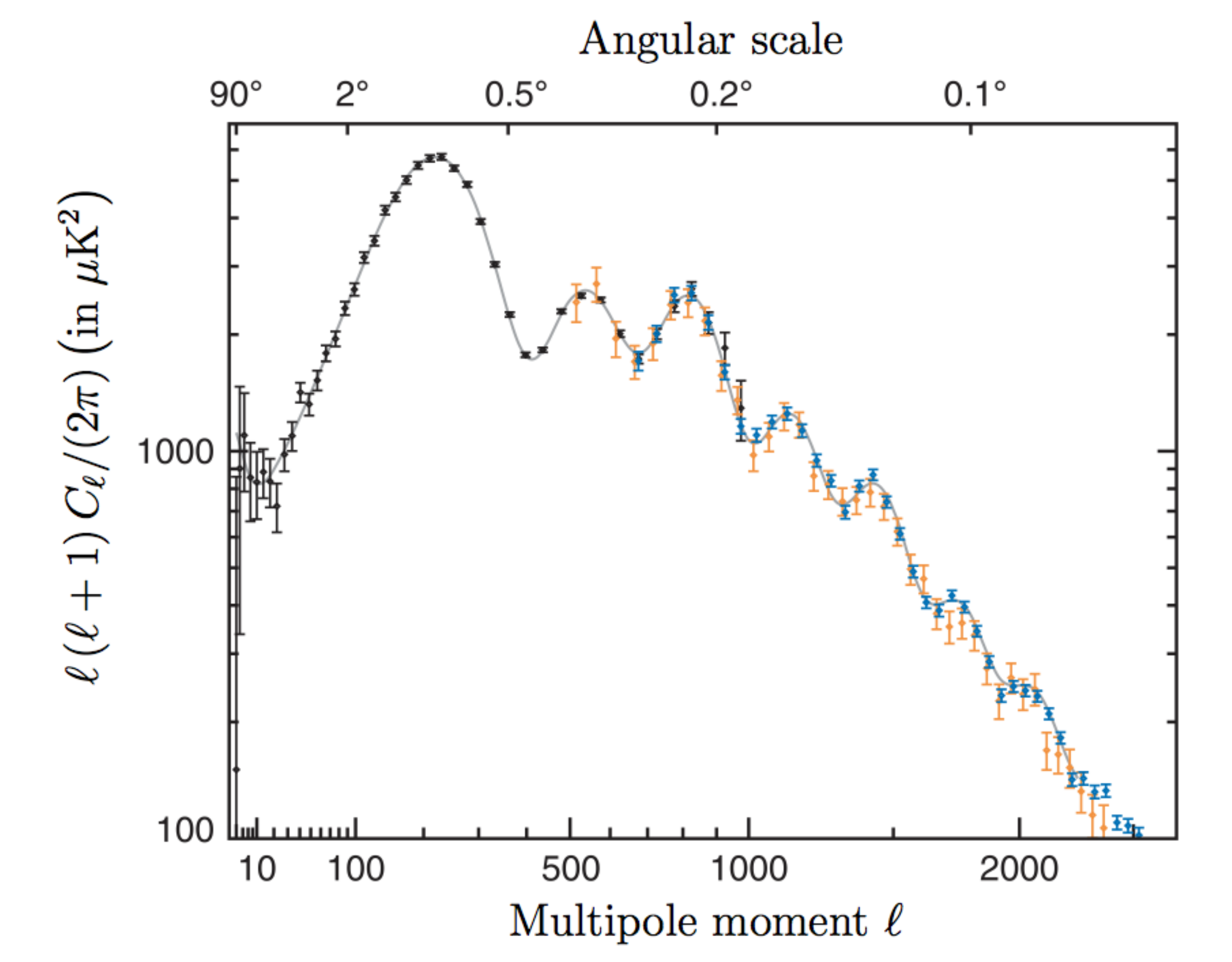}
\caption{Comparison between various measurements of the CMB
fluctuations, i.e. essentially the Fourier transform of the angular
power spectrum of the microwave light coming from the recombination
epoch (see other contributions in this volume). The figure shows the most
recent nine-year WMAP data, together with those coming from
SPT and ACT for the small scales. The standard $\Lambda$CDM
model shown here is merely coming from a fit with only the WMAP
data, which is then used to predict the higher multipole data: clearly
a quite precise and correct prediction!\newline
Figure taken from \cite{WMAP9} in which all references to other data are provided.}
\label{CMB}
\end{figure}

Now our cosmological model, precisely because of its successes, can be
scrutinized with exquisite attention to unveil any possible new mechanism
the we would not have thought about. This is how detailed examination
of specific objects (Type Ia SuperNov\ae) and their redshift distribution revealed
that the Universe appears to be currently accelerating (see however D.~Wiltshire's
contribution in this volume for an alternative understanding of the data), leading
to a new component, dubbed dark energy, among the various fluids pervading
the Universe. When added to the other components, it permits to fit all available
data, including large scale structure distribution, SuperNov\ae{} or the Cosmic
Microwave Background (CMB) fluctuations (see Fig.~\ref{CMB}).

There are many things lacking in this presentation, including not only the CMB
fluctuations themselves, but also its polarization and measurement thereof,
non gaussianities, baryonic acoustic oscillations, and many others. One special
point I would like to emphasize over however is that when all those data are
taken into account, the whole thing can serve not only to check the currently
accepted paradigms and models (an inflationary phase followed by radiation,
matter and cosmological constant dominations, the so-called $\Lambda$CDM
model\footnote{Meaning $\Lambda$ for the cosmological constant, and "Cold
Dark Matter", this phenomenological model describes with the minimal possible
number of parameters the current data.}), but also to explore alternative
possibilities: although the inflationary paradigms seems to provide satisfactory
explanations to most of the cosmological puzzles, it is not still proven beyond
reasonable doubt, and besides, it has a few problems of its own.

Inflation is based on well under control physics, i.e. GR and quantum/classical
scalar fields, it can be implemented in high energy theories such as Grand Unification
or string theories, and it makes predictions which have been experimentally
shown to be compatible with observations... why would we therefore like to find
any alternative at all? First of all, it does not really solve all the puzzles, as in
particular, the question of homogeneity, although admittedly alleviated, is by no
means solved. Moreover, an inflationary phase usually begins from a singularity,
or from a quantum gravity fluctuating phase, which is not understood at all. Related
to this is the fact that even the largest possible scales observed today, i.e. that
of the Hubble radius, must have inflated and expanded from a time where it was
actually smaller than the Planck length itself. Setting initial conditions there is,
to say the least, debatable. Finally, providing challengers is always a very good
way to test a theory, so inflation itself benefits from alternative models.

Most alternative to inflation present, in one way or another, a contracting
phase preceding the currently expanding one, to which it is related by means
of a bounce. This is not a new idea, as it was in fact suggested in the 1930's
by Tolman and Lema\^\i{}tre, i.e. much before any inflationary scenario was
even thought about. In course of time, bouncing scenarios were repeatedly
proposed, as discussed in Ref.~\cite{Novello08}. One might immediately
argue that this seems to create more problems than it solves, since in
particular it is very difficult to implement a bouncing phase in the framework
of GR; however, the bouncing model also addresses different issues. For
instance, there is of course no question of the primordial singularity, which is
avoided by definition! Moreover, the horizon (\ref{hor}) can easily be made
infinite if the initial time is sufficiently large and negative, i.e. in the limit
$t_\mathrm{ini}\to-\infty$. Flatness is also quite a natural achievement of 
the bounce, as I discussed earlier.

Now perturbation theory ought to be valid as well in a contracting background,
so basically, all I said before applies straightforwardly in such a new framework.
What needs be done then is to evolve similarly set vacuum initial conditions
in the contracting Universe all the way to the bounce and up to now. In general,
what happens is the following: whenever the relevant equation of the perturbations
takes the form (\ref{Sch}), the potential $U(\eta)$ can be more complicated
than that shown in Fig.~\ref{PotPert}, and in particular it often happens that
the term $k^2-U$ changes sign more than once or twice. As a result, the primordial
spectrum starts oscillating before it gets amplified again, and one expects
oscillations on top of the usual and expected oscillations. For the time being,
no observation has been made along these lines, but one can hope to see
those in the future, e.g. with Planck data.

Finally, I should say that the perturbation question is somehow an open one
in bouncing scenarios, and for many reasons. The first concerns for instance
the vector modes: as I said before, one usually neglects them as they decay
anyway with the expansion. Clearly, during contraction, one expects vector
modes to grow, and therefore they might pile up to produce unwanted non
linear vector-like objects, thus ruling out irremediably the corresponding
model. Therefore, one needs to check every model and initial condition setup,
although the situation is often quite unclear because without any specific
coupling with the matter fields, the vector modes are not dynamical, so
setting initial conditions for them is not feasible in any known natural way.
Scalar modes themselves can grow very large, but then comes the question
of gauge: is it absolutely clear that a large value for, say, the Bardeen
potential, means the theory becomes non linear? As a matter of fact, this
is yet undecided, and there are good arguments suggesting that providing
there exists a set of variables that behave perturbatively all through the
evolution of the Universe, then this set of variables should be used, at the
cost of breaking ``gauge invariance'', and the theory would still make sense.

Both inflationary and alternative models will probably be with us for still
quite a while, unless some (always possible) unexpected prediction or
observation comes in the way. In any case, we are living a very exiting
period, not only of the history of the Universe itself, but also in cosmology
where paradigm shifts are happening and new developments are proposed
at an ever increasing rate. With the advent of forthcoming data (Planck
of course, but also all the new proposals that just await actual construction),
it should not take long before new ideas come in the front stage... hopefully,
most of these notes will remain essentially valid.

\begin{theacknowledgments}
 I wish to acknowledge a superb school which was great
 at all possible levels, with very good lectures (hopefully
 including the present one!) during which
 I learned a lot, and enthusiastic students who made lecturing
 extremely enjoyable. Hence, my warms thanks go to the
 organizers Santiago Perez Bergliaffa and M\'ario Novello. I also
 want to thank my collaborators Nelson Pinto-Neto and J\'er\^ome
 Martin with whom I first came in contact with these topics. I also
 wish to thank J.~Martin, F.~A.~Teppa Pannia and S.Vitenti
 for careful reading of the manuscript.
\end{theacknowledgments}

\bibliographystyle{aipproc}


\end{document}